\documentclass[twocolumn,superscriptaddress]{revtex4-2}

\usepackage[T1]{fontenc}
\usepackage[utf8]{inputenc} 
\usepackage[english]{babel}
\usepackage{graphicx}
\usepackage{float}
\usepackage{caption}
\usepackage{subcaption}
\usepackage{amsmath, amssymb, amsfonts}
\usepackage{mathtools}
\usepackage{bm}
\usepackage{bbm} 
\usepackage{chemformula}
\usepackage{physics}
\usepackage{tabularx, booktabs}
\usepackage{xspace}
\usepackage{listings}
\usepackage{tikz}
\usetikzlibrary{arrows.meta, shapes}
\usepackage{gnuplottex}
\usepackage{placeins}
\usepackage{caption}
\usepackage[colorlinks=true, linkcolor=blue, citecolor=blue]{hyperref}
\usepackage{algorithm}
\usepackage[noend]{algpseudocode}
\usepackage{algorithm}
\usepackage{algpseudocode}

\usepackage{ulem}

\AtBeginEnvironment{bmatrix}{\setlength{\arraycolsep}{2.5pt}}

\makeatletter
\def\mathcolor#1#{\@mathcolor{#1}}
\def\@mathcolor#1#2#3{%
  \protect\leavevmode
  \begingroup
    \color#1{#2}#3%
  \endgroup
}
\makeatother

\newcommand{\manu}[1]{{\color{ForestGreen}#1}}
\newcommand{\inner}[2]{\left\langle #1 \middle\vert #2 \right\rangle}
\newcommand{\innerop}[3]{\left\langle #1 \middle\vert #2 \middle\vert #3 \right\rangle}
\newcommand{\undern}{\underline{n}}

\newcommand{\boldv}{\bf{v}}

\newcommand{\undernu}{\underline{\nu}}
\newcommand{\myexpval}[2]{\left\langle #1\right\rangle_{#2}}
\newcommand{\ie}{{\it i.e.}}

\newcommand{\be}{\begin{eqnarray}}
\newcommand{\ee}{\end{eqnarray}}
\newcommand{\myket}[1]{\left\vert #1\right\rangle}

\newcommand{\mybra}[1]{\left\langle #1\right\vert}

\newcommand{\rev}[1]{{{#1}}}

\DeclareMathOperator*{\argmin}{arg\,min}

\makeatletter
\def\@email#1#2{%
 \endgroup
 \patchcmd{\titleblock@produce}
  {\frontmatter@RRAPformat}
  {\frontmatter@RRAPformat{\produce@RRAP{*#1\href{mailto:#2}{#2}}}\frontmatter@RRAPformat}
  {}{}
}%
\makeatother
 
\begin{document}

\title[]{Local Potential Functional Embedding Theory of Molecular Systems: A Density-Functional Approach to Localized Orbital-Based Embedding}

\author{Wafa Makhlouf}
\affiliation{Laboratoire de Chimie Quantique, Institut de Chimie, Université de Strasbourg, 4 rue Blaise Pascal, 67000 Strasbourg, France}
\email{wmakhlouf@unistra.fr}

\author{Bruno Senjean}
\affiliation{ ICGM, Université de Montpellier, CNRS, ENSCM, Montpellier, France}
\date{\today}

\author{Emmanuel Fromager}
\affiliation{Laboratoire de Chimie Quantique, Institut de Chimie,
Université de Strasbourg, 4 rue Blaise Pascal, 67000 Strasbourg, France}

\begin{abstract}
Localized orbital-based quantum embedding, as originally formulated in the context of density matrix embedding theory (DMET), is revisited from the perspective of lattice density functional theory (DFT). An in-principle exact (in the sense of full configuration interaction) formulation of the theory, where the occupations of the localized orbitals play the role of the density, is derived for any (model or {\it ab initio}) electronic Hamiltonian. From this general formalism we deduce an exact relation between the local Hartree-exchange-correlation (Hxc) potential of the full-size Kohn--Sham (KS) lattice-like system and the embedding chemical potential that is adjusted on each embedded fragment, individually, such that both KS and embedding cluster systems reproduce the exact same local density. When well-identified density-functional approximations (that find their justification in the strongly correlated regime) are applied, a practical self-consistent local potential functional embedding theory (LPFET), where the local Hxc potential becomes the basic variable, naturally emerges from the theory. LPFET differs from previous density embedding approaches by its fragment-dependent embedding chemical potential expression, which is a simple functional of the Hxc potential. 
Numerical calculations on prototypical systems show the ability of such an ansatz to improve substantially the description of density profiles (localized orbitals occupation numbers in this context) in strongly correlated systems.     
\end{abstract}

\maketitle

\section{Introduction}

Quantum embedding theory~\cite{sun2016quantum,wasserman2020quantum,Evangelista_2025_Concluding,mejuto2024quantum} is a very active and promising research line in the field of electronic structure theory, in particular for treating strongly correlated molecular systems and materials. Even though the strategy for solving the electronic Schr\"{o}dinger equation is always the same (it can be summarized as ``divide and conquer''~\cite{marecat2023unitary}), there is no unique embedding theory, simply because the choice of the basic (few-electron reduced) quantity it can rely on is in fact arbitrary. A choice is ultimately made, thus leading to one embedding theory among many others, often motivated by either formal (mathematical) or practical advantages, or both. In this work we focus on density matrix embedding theory (DMET)~\cite{knizia2012density,
    knizia2013density,
    cances2025analysis,
    wouters2017five,
    wouters2016practical,
    chen2014intermediate,
    cui2020ground,
    kawano2020comparative,
    mukherjee2017simple,
    sandhoefer2016density,
    zheng2016ground,
    zheng2017stripe} and, more specifically, on density embedding theory (DET)~\cite{bulik2014density,bulik2014electron,fulde2017dealing,plat2020entanglement}. The two theories differ by the quantity that drives the self-consistent convergence of the embedding: DET only uses the diagonal part of the one-electron reduced density matrix (1RDM) for that purpose, hence its name. Since the seminal paper of Knizia and Chan~\cite{knizia2012density}, DMET has been intensively studied~\cite{tsuchimochi2015density,faulstich2022pure,cances2025analysis}, improved~\cite{fan2015cluster,
    gunst2017block,fertitta2019energy, wu2019projected,
    scott2021extending,
    zheng2017cluster,bulik2014density,
    tsuchimochi2015density, wu2020enhancing,
    negre2025new,
    nusspickel2022systematic,ma2024quantum}, and extended~\cite{chen2014intermediate,
    booth2015spectral,
    fertitta2018rigorous, kretchmer2018real,
    sriluckshmy2021fully,
    zheng2017cluster,tran2019using,
    mitra2021excited,
    sun2020finite,
    cernatic2024fragment,Booth24_PRL_Rigorous_Screened_Interactions}, not only for lattice models but also for {\it ab initio} systems~\cite{hermes2019multiconfigurational,
    hermes2020variational,
    cui2020efficient,
    haldar2023local,
    jangid2024core,
    cui2022systematic,
    ai2022efficient,
    ai2025density,
    ai2025exploring,
    lau2021regional,
    li2023multi,
    mitra2022periodic,
    mitra2023density,
    pham2019periodic,
    pham2018can,
    reinhard2019density}. It stimulated also new developments in alternative but related approaches like, for example, bootstrap embedding theory~\cite{welborn2016bootstrap,
    ricke2017performance,
    ye2019atom,
    ye2020bootstrap,
    tran2020bootstrap,
    tran2024bootstrap,
    meitei2023periodic} (together with extension to excited states~\cite{ye2021accurate} and Quantum computing~\cite{weisburn2025multiscale,
    erakovic2025high}), incremental embedding theory~\cite{ye2018incremental}, the (Ghost) Gutzwiller method~\cite{lanata2023derivation,
mejuto2024quantum}, the rotationally invariant slave boson approach~\cite{ayral2017dynamical,lee2019rotationally}, or the exact factorization of electronic wavefunctions~\cite{lacombe2020exactfac, requist2021fset}. The original formulation of DMET is in principle exact but not very useful as such, as it relies on the Schmidt decomposition of the exact many-body wavefunction~\cite{knizia2012density}, which is of course unknown, otherwise the embedding would be pointless. On the other hand, its standard implementation uses a single Slater determinant only, on which locally correlated 1RDM elements are mapped, in a completely frequency-independent setting~\cite{wouters2016practical}. This strategy can be seen as a drastic simplification of dynamical mean field theory (DMFT)~\cite{georges1992infdim, georges1996dmftreview, kotliar2004strongly, zgid2011dmftqc,Jiachen24_Interacting-Bath}, where the (frequency-dependent) local Green's function is the basic variable. Note that a different (also frequency-dependent) embedding theory can be formulated by using the self-energy as basic ingredient instead ~\cite{lan_generalized_2017}. Being a static approach (that uses, in addition, a very small quantum bath), DMET is a computationally cheap method that is applicable to both large molecules and materials. On a more fundamental level, it is still unclear, however, if its practical formulation (that is often presented as an approximation to the original derivation of DMET~\cite{knizia2012density}) can actually be made formally exact, in the spirit of Kohn--Sham (KS) DFT~\cite{kohn1965ks, hohenberg1964electron}. The motivation is not only theoretical. It is also practical. Indeed, by writing down such a theory, we would reconsider standard implementations of DMET (or DET) as approximations to some exact and well-defined functional(s), thus paving the way towards a systematic improvement of the method. It would also put recent developments in the field in a rigorous and broader perspective.\\ 
As our goal is to exactify standard DMET, \rev{our first concern should be} the 1RDM non-interacting (or mean-field) representability issue~\cite{faulstich2022pure}. Indeed, unlike correlated 1RDMs, the 1RDM of a Slater determinant is idempotent, \rev{thus making the mapping of the full (interacting) ground-state 1RDM onto a fictitious non-interacting system (that would also be in its ground state) impossible}. In this work, we simply bypass this problem by mapping the density only (\ie, the diagonal elements of the 1RDM), like in DET. 
\rev{We can reasonably assume that, like in regular KS-DFT, any exact (correlated) density profile can match that of a single Slater determinant. In this case, the latter} determinant is becoming a KS-like wavefunction, in the sense of DFT for lattices~\cite{gunnarsson1986density,schonhammer1995density,lima2002density,capelle2003density,Lima2003_density_functionals_luttinger_liquid}, the lattice sites being here the localized orbitals in which the 1RDM is represented. This connection with DFT has been exploited in the more recent self-consistent density embedding (SDE) approach~\cite{mordovina2019self} but not formulated as an exact theory, in the sense of full configuration interaction (FCI), at any level of fragmentation. To the best of our knowledge, such an exact theory has been derived (by one of the authors and coworkers~\cite{sekaran2022local}) only for the uniform one-dimensional Hubbard lattice model in the thermodynamic limit. Referred to as local potential functional embedding theory (LPFET), its extension to (finite and non-uniform) molecular systems is in principle not straightforward since each fragment has its own embedding cluster, as translational invariance cannot be exploited anymore. Most importantly, it is unclear how the embedding chemical potentials that are introduced within each cluster, in order to reproduce the local KS densities, are related, in an exact theory, to the (local) KS potential of the full-size lattice system.\\ 
The above questions are addressed in the following. We also discuss and exploit the practical implications of our findings. The paper is structured as follows. After reviewing briefly the standard practical formulation of DMET in Sec.~\ref{sec:DMET_review}, and exploiting in Sec.~\ref{eq:dmet_applied_to_KS_syst} its exactness in the particular case of a non-interacting (KS) system, we derive in Sec.~\ref{sec:exact_DET} the desired exact density-functional embedding theory. The key ingredient in this task will be the fragmentation of the analogue, in this context, to the Hohenberg--Kohn (HK) functional~\cite{hohenberg1964electron}. By introducing in Sec.~\ref{sec:LPFET_from_exact_dens_func_embedding_theory} well-defined density-functional approximations, that are all based on the assumption that electron correlation is strong and local, we obtain an LPFET that is now applicable to any system. A striking difference between LPFET and DET is the use of a fragment-dependent embedding chemical potential, in place of a global one. It emerges directly from the exact theory and is a simple functional of the local Hartree-exchange-correlation (Hxc) potential that defines the KS lattice system. Further insight is given in Sec.~\ref{sec:rationale_and_extensions_LPFET}. Following implementation and computational details in Sec.~\ref{sec:comput_details}, results obtained for prototypical six-site half-filled Hubbard models and a chain of six hydrogen atoms are shown and discussed in detail in Sec.~\ref{Results}. They demonstrate the ability of LPFET to capture more physics than DET in the strongly correlated regime. Conclusions and perspectives are finally given in Sec.~\ref{Con}.\\

\section{Theory} \label{Theory}

\subsection{A brief review of the DMET clusterization procedure} \label{sec:DMET_review}
Before introducing our embedding approach, it is helpful to briefly review the standard DMET clusterization procedure. In DMET, the full quantum system is divided into a smaller fragment of interest and an effective environment called the bath. The fragment and bath together define the embedding cluster, which is treated with a high-level method to capture the electronic properties of the fragment.
We start from the general second-quantized electronic Hamiltonian expression,
\be\label{eq:general_true_physical_Hamil}
\hat{H}=\hat{h}+\hat{U},
\ee
where the one- and two-electron contributions read 
\be
\hat{h}=\sum_{ij}h_{ij}\sum_{\sigma}\hat{c}_{i\sigma}^\dagger\hat{c}_{j\sigma}
\ee
and
\be\label{eq:general_two-electron_repulsion_op}
\hat{U}=\dfrac{1}{2}\sum_{ijkl}\sum_{\sigma,\sigma'}\langle ij\vert kl\rangle\hat{c}_{i\sigma}^\dagger\hat{c}_{j\sigma'}^\dagger\hat{c}_{l\sigma'}\hat{c}_{k\sigma},
\ee
respectively. In the present context, the one-electron states $\ket{\chi_{i\sigma}} \equiv \hat{c}^{\dagger}_{i\sigma} \ket{\text{vac}}$ are generated from an orthonormal localized orbital basis $\left\{\chi_i\right\}$. The localized character of the latter orbitals makes the embedding of strongly correlated electrons (that is discussed in the following) useful and efficient. Interestingly, the formulation of the embedding is general, \ie, it does not depend on the orbital localization scheme. In addition, it applies to both regular {\it ab initio} quantum chemical Hamiltonians and lattice model Hamiltonians, which are popular in condensed matter physics~\cite{DFT_ModelHamiltonians}. In the latter case, the index $i$ simply refers to an atomic {\it site} (terminology that we will use throughout the paper) while the one-electron (kinetic and nuclear attraction) $h_{ij}$ and two-electron repulsion $\langle ij\vert kl\rangle$ integrals are simply used as parameters.\\     

For the sake of simplicity and clarity, we focus in this work on single-orbital embeddings (and their formal exactification within the KS formalism) but our strategy can be extended to multiple-orbital embeddings along the exact same lines. In the standard DMET setting~\cite{knizia2012density}, a given (often referred to as fragment) orbital $\chi_i$ of interest is embedded into a (spin-restricted, in the present case) quantum bath that consists of a single bath orbital (the bath contains as many orbitals as the fragment when multiple fragment orbitals are embedded~\cite{sekaran2023unified}). The latter is constructed as follows~\cite{sekaran2023unified},        
\be\label{eq:bath_from_1RDM}
\ket{b^{(i)}}=\sum_{j\neq i}\frac{\gamma^\sigma_{ij}}{\sqrt{\sum_{k\neq i}\abs{\gamma^\sigma_{ki}}^2}}\ket{\chi_j},
\ee
where the elements $\gamma^\sigma_{ij}=\langle\hat{c}_{i\sigma}^\dagger\hat{c}_{j\sigma}\rangle$ of the 1RDM (simply referred to as {\it density matrix} in the following) are evaluated from a single Slater determinant that is computed (self-consistently) as an approximation to the {\it full}-system many-body wavefunction. In this context, self-consistency is induced by density (matrix) mapping constraints, hence the name of the theory. Often referred to as the ``low-level'' part of the DMET calculation, we will show later in Sec.~\ref{sec:exact_DET} that, if we restrict ourselves to a density (diagonal of the density matrix) mapping, it can actually be exactified by invoking the KS formalism of DFT. Strictly speaking, DMET becomes a DET~\cite{bulik2014density} in this case. In the so-called interacting bath (IB) formulation of DMET~\cite{wouters2016practical}, which is considered in this work unless otherwise indicated, the full Hamiltonian $\hat{H}$ is projected onto the ``active'' space $\left\{\Phi_{\alpha}^{(i)}\right\}$ corresponding to the half-filled (two-electron, in the present single-orbital embedding case) ``fragment+bath'' embedding cluster in the presence of the remaining (core) electrons, which are described by the single Slater determinant $\Phi^{(i)}_{\rm core}$, \ie,
\be
\hat{H}\rightarrow \hat{P}^{(i)}\hat{H}\hat{P}^{(i)},
\ee
where
\be\label{eq:projector_op_onto_embedding_cluster}
\hat{P}^{(i)}=\sum_{\alpha}\myket{\Phi^{(i)}_{\rm core}\Phi_{\alpha}^{(i)}}\mybra{\Phi^{(i)}_{\rm core}\Phi_{\alpha}^{(i)}}.
\ee
Note that both bath and core orthonormal orbitals can be generated straightforwardly by applying a Householder transformation to the density matrix~\cite{sekaran2021householder,yalouz2022quantum,sekaran2023unified}. The final embedding cluster Hamiltonian, that can be solved exactly (or accurately, in the multiple-orbital embedding case), is obtained by introducing a global ($i$-independent) chemical potential on the {\it embedded} fragment orbital $\chi_i$ (that is then referred to as {\it impurity}, following the terminology of DMFT~\cite{georges1992infdim, georges1996dmftreview, kotliar2004strongly, zgid2011dmftqc,Jiachen24_Interacting-Bath}),
\be\label{eq:_global_mu_introduced_DMET}
\hat{P}^{(i)}\hat{H}\hat{P}^{(i)}\rightarrow \hat{\mathcal{H}}^{(i)}=\hat{P}^{(i)}\hat{H}\hat{P}^{(i)}-\mu_{\rm glob}\hat{n}_i,
\ee
where $\hat{n}_i=\sum_{\sigma=\uparrow,\downarrow}\hat{n}_{i\sigma}$ (with $\hat{n}_{i\sigma}=\hat{c}_{i\sigma}^\dagger\hat{c}_{i\sigma}$) denotes the impurity occupation operator, such that, when combining all the embedding clusters, the correct total number $N$ of electrons in the original full system is obtained:     
\be
\sum_i\left\langle\hat{n}_i\right\rangle_{\hat{\mathcal{H}}^{(i)}}\overset{!}{=}N.
\ee
It then becomes possible to approximate local properties such as the fragment single and double occupations or off-diagonal elements of the density matrix as follows,
\begin{subequations}\label{approximate_RDMs_from_cluster}
\begin{align}
\left\langle\hat{n}_i\right\rangle_{\hat{H}}&\approx \left\langle\hat{n}_i\right\rangle_{\hat{\mathcal{H}}^{(i)}},
\\
\left\langle\hat{n}_{i\uparrow}\hat{n}_{i\downarrow}\right\rangle_{\hat{H}}
&\approx \left\langle\hat{n}_{i\uparrow}\hat{n}_{i\downarrow}\right\rangle_{\hat{\mathcal{H}}^{(i)}},
\\
\left\langle\hat{c}_{i\sigma}^\dagger\hat{c}_{j\sigma}\right\rangle_{\hat{H}}&\underset{j\neq i}{\approx}
\left\langle\hat{c}_{i\sigma}^\dagger\hat{c}_{j\sigma}\right\rangle_{\hat{\mathcal{H}}^{(i)}}.
\end{align}
\end{subequations}
The proper reconstruction of the total $N$-electron energy from the embedding clusters is a non-trivial and still challenging task (in particular because of the associated $N$-representability problem~\cite{nusspickel2023effective}) that is not central in the present work. It will be briefly discussed later in Sec.~\ref{Hubbard_ring}.\\

At this point it is important to stress that, in the (unphysical) case of non-interacting electrons ($\hat{U}\equiv 0$), the embedding procedure becomes {\it exact}, 
with no need for a global chemical potential ($\mu_{\rm glob}=0$), meaning that Eqs.~(\ref{approximate_RDMs_from_cluster}) are not approximations anymore~\cite{sekaran2023unified}. In fact, this statement holds for mean-field electrons as well, since they are also described by an idempotent density matrix~\cite{sekaran2023unified}. The above observation is our main motivation for invoking the KS-DFT formalism in the formal exactification of the embedding procedure, as explained in further detail in the following. 

\subsection{Applying DMET to KS lattice DFT}\label{eq:dmet_applied_to_KS_syst}

By analogy with DFT for lattice models~\cite{DFT_ModelHamiltonians,Coe_2015_Uniqueness,Fromager_2015,Franca2018_Testing,Penz_2021_Density,Penz2023geometryof,Penz_2024_Geometrical}, we rewrite the true physical Hamiltonian \rev{of Eq.~(\ref{eq:general_true_physical_Hamil})} as follows,
\begin{equation}
    \hat{H}= \hat{T}+\hat{U}+\hat{V}^{\rm ext},
\end{equation}
where \rev{$\hat{U}$ denotes the electronic repulsion operator (see Eq.~(\ref{eq:general_two-electron_repulsion_op})), the ``off-diagonal'' contribution $\hat{T}:=\sum_{i\neq j}h_{ij}\sum_{\sigma}\hat{c}_{i\sigma}^\dagger\hat{c}_{j\sigma}$ to the one-electron part of the Hamiltonian ($\hat{h}$ in Eq.~(\ref{eq:general_true_physical_Hamil})) plays a similar role as the kinetic (hopping) energy operator in a lattice model (the sites being the localized orbitals in the present context), while the ``diagonal'' contribution to $\hat{h}$, denoted $\hat{V}^{\rm ext}:=\sum_i v_i^{\rm ext}\hat{n}_i$, with $v_i^{\rm ext}=h_{ii}$, plays the role of the external potential that controls the electronic density in the lattice, \ie, the occupation of the sites}. On that basis, the exact $N$-electron ground-state energy of $\hat{H}$ (``exact'' meaning here FCI) can be evaluated \`{a} la KS~\cite{kohn1965ks, hohenberg1964electron} as follows,
\begin{equation}\label{eq:KSDFT_var_principle}
    E=\min_{\Phi\rightarrow N}\left\{\innerop{\Phi}{\hat{T}+\hat{V}^{\rm ext}}{\Phi}+E_{\rm Hxc}\left(\undern^{\Phi}\right)\right\},
\end{equation}
where \rev{$\Phi$ is a trial $N$-electron single-determinant wavefunction}, $\undern^{\Phi}\equiv \left\{\innerop{\Phi}{\hat{n}_i}{\Phi}\right\}$ and, for any $N$-electron density profile 
\be
\undern\equiv \left\{n_i\right\},\;{\sum_i n_i=N},
\ee
the Hxc density-functional energy reads in this context
\rev{
\be\label{eq:Hxc_ener_exp_from_F}
E_{\rm Hxc}\left(\undern\right)&=F(\undern)-T_{\rm s}(\undern),
\ee
\be\label{eq:HK_func_def}
F(\undern)
=\min_{\Psi\rightarrow \undern}\innerop{\Psi}{\hat{T}+\hat{U}}{\Psi}
=\innerop{\Psi(\undern)}{\hat{T}+\hat{U}}{\Psi(\undern)}
\nonumber \\
\ee
and
\be\label{eq:Ts_func_def}
\begin{split}
T_{\rm s}(\undern)&=\min_{\Psi\rightarrow \undern}\innerop{\Psi}{\hat{T}}{\Psi}
&=\innerop{\Phi(\undern)}{\hat{T}}{\Phi(\undern)}
\end{split}
\ee
being the analogs for lattices of the fully-interacting Levy--Lieb~\cite{levy1979universal,LFTransform-Lieb} and non-interacting kinetic energy density functionals, respectively, with the density constraint $\Psi\rightarrow \undern$ in both minimizations meaning $\undern^\Psi\equiv \left\{\innerop{\Psi}{\hat{n}_i}{\Psi}\right\}=\undern$. The minimizers $\Psi(\undern)$ and $\Phi(\undern)$ of Eqs.~(\ref{eq:HK_func_def}) and (\ref{eq:Ts_func_def}) are the interacting and fictitious KS non-interacting ground states [in the latter case, $T_{\rm s}(\undern)$ is the kinetic energy of the density-functional KS determinant $\Phi(\undern)$], respectively, both having the exact same density profile $\undern$. For completeness, let us recall that, by analogy with regular (real-space) KS-DFT, the exact Hx density-functional energy reads $E_{\rm Hx}\left(\undern\right)=\innerop{\Phi(\undern)}{\hat{U}}{\Phi(\undern)}$ so that, according to Eqs.~(\ref{eq:Hxc_ener_exp_from_F})-(\ref{eq:Ts_func_def}), the exact correlation density-functional energy $E_{\rm c}\left(\undern\right)=E_{\rm Hxc}\left(\undern\right)-E_{\rm Hx}\left(\undern\right)$ can be expressed as follows,
\begin{eqnarray}
E_{\rm c}\left(\undern\right)&=&\innerop{\Psi(\undern)}{\hat{T}+\hat{U}}{\Psi(\undern)}-
\innerop{\Phi(\undern)}{\hat{T}+\hat{U}}{\Phi(\undern)}
\nonumber
\\ 
&=&U_{\rm c}\left(\undern\right)+T_{\rm c}\left(\undern\right),
\end{eqnarray}
where $U_{\rm c}\left(\undern\right)=\innerop{\Psi(\undern)}{\hat{U}}{\Psi(\undern)}-
\innerop{\Phi(\undern)}{\hat{U}}{\Phi(\undern)}$ is the potential correlation energy contribution while $T_{\rm c}\left(\undern\right)=\innerop{\Psi(\undern)}{\hat{T}}{\Psi(\undern)}-
\innerop{\Phi(\undern)}{\hat{T}}{\Phi(\undern)}$ is the kinetic part of the total correlation energy (note that the same kinetic energy operator $\hat{T}$ applies to both interacting and non-interacting wavefunctions). A similar decomposition will be invoked later in Sec.~\ref{sec:exact_DET} when deriving an exact density-functional embedding theory.\\ 
}

A key feature of KS-DFT that will become central later when turning to quantum embedding theory is the fact that the minimizing KS determinant $\Phi^{\rm KS}$ in Eq.~(\ref{eq:KSDFT_var_principle}) reproduces the density $\undern^{\Psi_0}$ of the exact physical ground state,
\be
\Psi_0=\argmin_{\Psi\rightarrow N}\innerop{\Psi}{\hat{H}}{\Psi}.
\ee
Moreover, $\Phi^{\rm KS}$ is the ground state of the non-interacting density-functional KS Hamiltonian, \ie, 
\be\label{eq:KS_VP}
\Phi^{\rm KS}=\Phi\left(\undern^{\Psi_0}\right)=\argmin_{\Phi\rightarrow N}\innerop{\Phi}{\hat{T}+\sum_iv^{\rm s}_i\hat{n}_i}{\Phi},
\ee
where the value of the KS potential on site $i$ reads
\be\label{eq:exact_KS_pot_on_site_i}
v^{\rm s}_i:=v^{\rm s}_i\left(\undern^{\Psi_0}\right)=v_i^{\rm ext}+v_i^{\rm Hxc}.
\ee
The exact Hxc potential can be expressed formally as follows, 
\be\label{eq:Hxc_pot_func_deriv}
v_i^{\rm Hxc}=\left.\dfrac{\partial E_{\rm Hxc}({\bf n})}{\partial n_i}\right|_{{\bf n}=\undern^{\Psi_0}},
\ee
where, in the more general density profile ${\bf n}\equiv \left\{n_i\right\}$, each orbital occupation $n_i$ can vary independently of the others, thus allowing for infinitesimal deviations from purely $N$-electron density profiles. This subtle point will become important later in Sec.~\ref{sec:rationale_and_extensions_LPFET} \rev{(see also Appendix~\ref{app:alternative_deriv_LPFET})} when rationalizing the LPFET approach.\\

At this point we should stress that, unlike in standard (real space) DFT, we are not aiming at developing or using density-functional approximations (DFAs) to $E_{\rm Hxc}({\bf n})$ that would depend explicitly on $\undern$. This has been done only in the very specific case of lattice model Hamiltonians where the kinetic and two-electron repulsion energies are controlled by a few parameters~\cite{DFT_ModelHamiltonians}. Instead, we want to approach $E_{\rm Hxc}\left(\undern^{\Psi_0}\right)$ or, equivalently (see Eq.~(\ref{eq:Hxc_ener_exp_from_F})), $F\left(\undern^{\Psi_0}\right)$, from embedding clusters designed \`{a} la DMET, where the reference ``low-level'' full-size calculation is the lattice KS-DFT one of Eq.~(\ref{eq:KS_VP}). 
In order to formally exactify the approach, we need an alternative (to the KS one in Eq.~(\ref{eq:Hxc_ener_exp_from_F})) density-functional decomposition of $F(\undern)$. We also need a clearer connection between the global chemical potential of DMET (see Eq.~(\ref{eq:_global_mu_introduced_DMET})) and all the components $\left\{v_i^{\rm Hxc}\right\}$ of the Hxc potential. In particular, in the general case of a non-uniform density profile, we can question the relevance of a global chemical potential versus a cluster-dependent one.\\ 

The above questions will be addressed in detail in the next section, thus generalizing (and extending to quantum chemistry) a previous work on the thermodynamic limit of the uniform one-dimensional Hubbard model~\cite{sekaran2022local}. As a warm-up, let us apply DMET to the KS system. For that purpose we decompose the (lattice) kinetic energy operator in terms of per-site contributions as follows,    
\be
\hat{T}= \sum_i\hat{t}^{(i)},
\ee
where
\be\label{eq:per-site_kinetic_ener_op_for_emb}
\hat{t}^{(i)}=\sum_{j>i}h_{ij}\sum_{\sigma}\left(\hat{c}_{i\sigma}^\dagger\hat{c}_{j\sigma}+\hat{c}_{j\sigma}^\dagger\hat{c}_{i\sigma}\right).
\ee
Since the KS determinant is a functional of the $N$-electron density profile $\undern$, its density matrix and, consequently (see Eq.~(\ref{eq:bath_from_1RDM})), the bath associated to the fragment orbital $\chi_i$ are functionals of $\undern$. According to Sec.~\ref{sec:DMET_review} and Eq.~(\ref{eq:KS_VP}), the resulting non-interacting KS density-functional embedding cluster Hamiltonian simply reads
\be\label{eq:KS_cluster_Hamil_full_proj}
\hat{\mathcal{H}}_{\rm s}^{(i)}(\undern)=\hat{P}^{(i)}(\undern)\left(\hat{T}+\sum_k v^{\rm s}_k(\undern)\hat{n}_k\right)\hat{P}^{(i)}(\undern).
\ee
At this point it is important to realize that the projection of the many-electron KS potential operator $\hat{V}^{\rm s}(\undern)=\sum_k v^{\rm s}_k(\undern)\hat{n}_k$ onto a given cluster $i$ (second term on the right-hand side of Eq.~(\ref{eq:KS_cluster_Hamil_full_proj})) can be substantially simplified. As similar simplifications will become crucial in the general interacting case (see Sec.~\ref{sec:exact_DET}), we provide below the important steps that lead to the final practical expression of Eq.~(\ref{eq:non_int_KS_cluster_dens_func_Hamil_all_on_imp}). For that purpose, we introduce the spin-free one-electron version of the KS potential operator, 
\be\label{eq:one-elec_KS_dens_func_pot_ket_bra}
\hat{v}^{\rm s}(\undern)=\sum_k v^{\rm s}_k(\undern)\myket{\chi_k}\mybra{\chi_k},
\ee
and rewrite the second term on the right-hand side of Eq.~(\ref{eq:KS_cluster_Hamil_full_proj}) in the core (see Eq.~(\ref{eq:projector_op_onto_embedding_cluster})), impurity, and bath orbitals representation, which gives 
\be\label{eq:detailed_deriv_dens_func_KS_cluster_Hamilto}
\begin{split}
\hat{P}^{(i)}(\undern)\hat{V}^{\rm s}(\undern)\hat{P}^{(i)}(\undern)
&\equiv\innerop{\Phi^{(i)}_{\rm core}(\undern)}{\hat{V}^{\rm s}(\undern)}{\Phi^{(i)}_{\rm core}(\undern)}
\\
&+\innerop{\chi_i}{\hat{v}^{\rm s}(\undern)}{\chi_i}\hat{n}_i \\
&
+\innerop{b^{(i)}(\undern)}{\hat{v}^{\rm s}(\undern)}{b^{(i)}(\undern)}\hat{n}_{b^{(i)}(\undern)},
\end{split}
\ee
where the contribution of the core (first term on the right-hand side of Eq.~(\ref{eq:detailed_deriv_dens_func_KS_cluster_Hamilto})) is a {\it scalar}. Note also the absence of impurity-bath coupling terms in the above equation, due to the locality of the KS potential in the original lattice representation and the orthogonality of the bath to the impurity (see Eqs.~(\ref{eq:bath_from_1RDM}) and (\ref{eq:one-elec_KS_dens_func_pot_ket_bra})):
\be
\innerop{\chi_i}{\hat{v}^{\rm s}(\undern)}{b^{(i)}(\undern)}
=v^{\rm s}_i(\undern)\inner{\chi_i}{b^{(i)}(\undern)}=0.
\ee
Finally, if we introduce the operator $\hat{N}^{(i)}(\undern)=\hat{n}_i+\hat{n}_{b^{(i)}(\undern)}$ that counts the total number of electrons in the cluster, Eq.~(\ref{eq:detailed_deriv_dens_func_KS_cluster_Hamilto}) becomes
\be\label{eq:detailed_deriv_dens_func_KS_cluster_Hamilto_step2}
\begin{split}
\hat{P}^{(i)}(\undern)\hat{V}^{\rm s}(\undern)\hat{P}^{(i)}(\undern)
\equiv
&\Big(\innerop{\chi_i}{\hat{v}^{\rm s}(\undern)}{\chi_i} \\
&-\innerop{b^{(i)}(\undern)}{\hat{v}^{\rm s}(\undern)}{b^{(i)}(\undern)}\Big)\hat{n}_i
\\
&
+\innerop{\Phi^{(i)}_{\rm core}(\undern)}{\hat{V}^{\rm s}(\undern)}{\Phi^{(i)}_{\rm core}(\undern)}
\\
&+\innerop{b^{(i)}(\undern)}{\hat{v}^{\rm s}(\undern)}{b^{(i)}(\undern)}\hat{N}^{(i)}(\undern).
\end{split}
\ee
Since the cluster is closed and contains two electrons, as explained in Sec.~\ref{sec:DMET_review}, $\hat{N}^{(i)}(\undern)\equiv 2$ can in fact be replaced by a {\it scalar} (we only look at the expression of the operators after projection onto the cluster). As a consequence, the last two terms on the right-hand side of Eq.~(\ref{eq:detailed_deriv_dens_func_KS_cluster_Hamilto_step2}) have no impact on the cluster's ground-state wavefunction and, therefore, can be removed. With the following additional simplifications (see Eqs.~(\ref{eq:bath_from_1RDM}) and (\ref{eq:one-elec_KS_dens_func_pot_ket_bra})), $\innerop{\chi_i}{\hat{v}^{\rm s}(\undern)}{\chi_i}=v^{\rm s}_i(\undern)$ and $\innerop{b^{(i)}(\undern)}{\hat{v}^{\rm s}(\undern)}{b^{(i)}(\undern)}=\sum_{k\neq i} v^{\rm s}_k(\undern)\inner{b^{(i)}(\undern)}{\chi_k}^2$, we obtain our final simplified expression for the KS embedding cluster's Hamiltonian of Eq.~(\ref{eq:KS_cluster_Hamil_full_proj}),
\be
\label{eq:non_int_KS_cluster_dens_func_Hamil_all_on_imp}
\begin{split}
\hat{\mathcal{H}}_{\rm s}^{(i)}(\undern)
\equiv & \hat{P}^{(i)}(\undern)\hat{T}\hat{P}^{(i)}(\undern)
\\
&
+\left(v^{\rm s}_i(\undern)-\sum_{k\neq i}\inner{b^{(i)}(\undern)}{\chi_k}^2v^{\rm s}_k(\undern)\right)\hat{n}_i,
\end{split}
\ee
where, in the light of Eq.~(\ref{eq:detailed_deriv_dens_func_KS_cluster_Hamilto}), it looks as if the KS potential value on the bath site $\innerop{b^{(i)}(\undern)}{\hat{v}^{\rm s}(\undern)}{b^{(i)}(\undern)}$ has been transferred by a ``constant'' shift (\ie, the same potential value has been subtracted on both cluster sites) to the impurity site $\chi_i$.
\\

By construction, the (two-electron, in the present single orbital embedding) ground state of the KS embedding cluster is determined as follows,
\be\label{eq:KS_dens_func_cluster_wf_def}
\Phi^{(i)}(\undern)=\argmin_{\Phi\rightarrow 2}\innerop{\Phi}{\hat{\mathcal{H}}_{\rm s}^{(i)}(\undern)}{\Phi}.
\ee
It reproduces the local density exactly, \ie,
\be\label{eq:local_dens_mapping_KS}
n_i=\innerop{\Phi(\undern)}{\hat{n}_i}{\Phi(\undern)}=\innerop{\Phi^{(i)}(\undern)}{\hat{n}_i}{\Phi^{(i)}(\undern)},
\ee
as well as the per-site non-interacting density-functional kinetic energy
\begin{eqnarray}
\innerop{\Phi(\undern)}{\hat{t}^{(i)}}{\Phi(\undern)}
&=&2\sum_{j>i}h_{ij}\sum_{\sigma}\innerop{\Phi(\undern)}{\hat{c}_{i\sigma}^\dagger\hat{c}_{j\sigma}}{\Phi(\undern)} \nonumber
\\
&=&2\sum_{j>i}h_{ij}\sum_{\sigma}\innerop{\Phi^{(i)}(\undern)}{\hat{c}_{i\sigma}^\dagger\hat{c}_{j\sigma}}{\Phi^{(i)}(\undern)} \nonumber
\\
&=&\innerop{\Phi^{(i)}(\undern)}{\hat{t}^{(i)}}{\Phi^{(i)}(\undern)},
\end{eqnarray}
which leads to the following exact evaluation, by clusterization, of the total non-interacting kinetic energy functional:  \be
T_{\rm s}(\undern)=\sum_i\innerop{\Phi^{(i)}(\undern)}{\hat{t}^{(i)}}{\Phi^{(i)}(\undern)}.
\ee
Consequently (see Eq.~(\ref{eq:Hxc_ener_exp_from_F})), the analog for lattices of the ``universal'' HK functional can be rewritten equivalently as follows, 
\be\label{eq:KS_decomp_from_clusters}
F(\undern)=\sum_i\innerop{\Phi^{(i)}(\undern)}{\hat{t}^{(i)}}{\Phi^{(i)}(\undern)}+E_{\rm Hxc}(\undern).
\ee
For the sake of clarity, note that the constraint $``\Phi\rightarrow 2"$ in Eq.~(\ref{eq:KS_dens_func_cluster_wf_def}) means that only the cluster part of the wavefunction is determined variationally while the core part is frozen and described by the single density-functional determinant $\Phi^{(i)}_{\rm core}(\undern)$. The exact same comment will apply later in Eq.~(\ref{eq:cluster_dens_func_wf_def}).



\subsection{Exact density embedding theory}\label{sec:exact_DET}

We now consider the following (in-principle exact) alternative density-functional clusterized expression,
\be\label{eq:clusterized_HK_func_exp}
F(\undern)=\sum_i\innerop{\Psi^{(i)}(\undern)}{\hat{t}^{(i)} + \hat{U}^{(i)}}{\Psi^{(i)}(\undern)}+\bar{E}_{\rm Hxc}(\undern), \nonumber \\
\ee
where $\hat{U}^{(i)}\equiv \inner{ii}{ii}\hat{n}_{i\uparrow}\hat{n}_{i\downarrow}$ denotes the interaction on the impurity $\chi_i$ and, following the DMET (IB) construction of Sec.~\ref{sec:DMET_review}, the density-functional embedding cluster wavefunction is defined as follows,
\be\label{eq:cluster_dens_func_wf_def}
\Psi^{(i)}(\undern)=\argmin_{\Psi\rightarrow 2}\innerop{\Psi}{\hat{\mathcal{H}}^{(i)}(\undern)}{\Psi},
\ee
where the density-functional embedding cluster Hamiltonian reads
\be\label{eq:dens_func_cluster_Hamil_main_text}
\hat{\mathcal{H}}^{(i)}(\undern)=\hat{P}^{(i)}(\undern)\hat{H}(\undern)\hat{P}^{(i)}(\undern)-\mu^{(i)}_{\rm imp}(\undern)\hat{n}_i,
\ee
$\hat{P}^{(i)}(\undern)$ being the density-functional onto-the-cluster projection operator introduced in Sec.~\ref{eq:dmet_applied_to_KS_syst} (it is still constructed from the full-size density-functional KS determinant $\Phi(\undern)$) and $\hat{H}(\undern)=\hat{T}+\hat{U}+\sum_i(v^{\rm s}_i(\undern)-v^{\rm Hxc}_i(\undern))\hat{n}_i$ denotes the fully-interacting and full-size density-functional Hamiltonian. The ({\it a priori} $i$-dependent) impurity chemical potential $\mu^{(i)}_{\rm imp}(\undern)$ has been introduced in Eq.~(\ref{eq:dens_func_cluster_Hamil_main_text}) in order to ensure that, like the KS system, the (IB) embedding cluster reproduces the local density exactly:   
\be\label{eq:exact_dens_constraint_cluster}
\innerop{\Psi^{(i)}(\undern)}{\hat{n}_i}{\Psi^{(i)}(\undern)}=n_i=\innerop{\Phi^{(i)}(\undern)}{\hat{n}_i}{\Phi^{(i)}(\undern)}.
\ee
Let us stress for clarity that, in this context, the terminology ``chemical potential'' applies to the embedded impurity alone, an open quantum system (whose occupation has to be determined, ultimately) that belongs to a closed embedding cluster. While the interactions beyond each fragment (a single orbital in the present formulation) are approximately described, by projection, within each cluster and therefore transferred to the embedding wavefunctions, the complementary Hxc functional $\bar{E}_{\rm Hxc}(\undern)$ in Eq.~(\ref{eq:clusterized_HK_func_exp}) enables to recover, in principle exactly, the missing contributions to the electronic repulsion energy. Its evaluation in practice, as an implicit functional of the density, will be discussed further in Sec.\ref{Results}.\\

In order to relate the impurity chemical potentials to the Hxc potential, let us (arbitrarily) choose one site, that we refer to as {\it residual site} and denote $r$, where the local density expression,
\be\label{eq:dens_residual_site}
n_r=N-\sum_{j\neq r}n_j,
\ee
ensures that the total number of electrons in the full system is fixed and equal to the integer number $N$ when density-functional derivatives are evaluated (our final conclusions should obviously not depend on this choice). As shown in the Appendix~\ref{appendix:relations_chem_pots_partial-int_bath} (see Eq.~(\ref{eqapp:general_imp_chem_pots_relation_part-IB})), all impurity chemical potentials can be expressed in terms of that on the residual site, as follows,  
\be\label{eq:exact_imp_chem_pots_relations}
\begin{split}
&-\mu^{(j)}_{\rm imp}(\undern)
\underset{j\neq r}{=}-\mu^{(r)}_{\rm imp}(\undern)\\
&+\sum_k\left(\inner{b^{(r)}(\undern)}{\chi_k}^2-\inner{b^{(j)}(\undern)}{\chi_k}^2\right)v_k^{\rm Hxc}(\undern)
\\
&+\dfrac{\partial \bar{E}_{\rm Hxc}(\undern)}{\partial n_j}
-\left.\dfrac{\partial \mathcal{U}_{\rm Hxc}(\undernu,\undern)}{\partial n_j}\right|_{\undernu=\undern}
-\left.\dfrac{\partial \tau_{\rm c}(\undernu,\undern)}{\partial n_j}\right|_{\undernu=\undern},
\end{split}
\ee
thus establishing the desired exact density-functional relation between them and the Hxc potential. This is our first key result. As readily seen from Eq.~(\ref{eq:exact_imp_chem_pots_relations}), which generalizes the derivation made in Ref.~\cite{sekaran2022local} for the uniform Hubbard model in the thermodynamic limit, two density bi-functionals, in addition to the complementary one $\bar{E}_{\rm Hxc}(\undern)$ introduced in Eq.~(\ref{eq:clusterized_HK_func_exp}), are in principle needed. The first one,
\be
\tau_{\rm c}(\undernu,\undern)=\sum_i\tau_{\rm c}^{(i)}(\undernu,\undern),
\ee
where
\be\label{eq:def_clusterization_kinetic_corr_ener_bifunc}
\begin{split}
\tau_{\rm c}^{(i)}(\undernu,\undern)&=\innerop{\Psi^{(i)}(\undern)}{\hat{\tau}^{(i)}(\undernu)}{\Psi^{(i)}(\undern)}
\\
&\quad
-\innerop{\Phi^{(i)}(\undern)}{\hat{\tau}^{(i)}(\undernu)}{\Phi^{(i)}(\undern)}
\end{split}
\ee
and 
\be\label{eq:tau_op_def}
\hat{\tau}^{(i)}(\undernu)=\hat{P}^{(i)}(\undernu)\hat{T}\hat{P}^{(i)}(\undernu)-\hat{t}^{(i)},
\ee
can be interpreted as a clusterization kinetic correlation energy functional, as shown in Appendix~\ref{appendix:clusterization_kin_corr_energy}. The second one, 
\be
\mathcal{U}_{\rm Hxc}(\undernu,\undern)=\sum_i \mathcal{U}^{(i)}_{\rm Hxc}(\undernu,\undern),
\ee
where
\be\label{eq:beyond_imp_within_cluster_Hxc_pot_ener_bi_func}
\begin{split}
&
\mathcal{U}^{(i)}_{\rm Hxc}(\undernu,\undern)
=
\innerop{\Psi^{(i)}(\undern)}{\hat{P}^{(i)}(\undernu)\left(\hat{U}-\hat{U}^{(i)}\right)\hat{P}^{(i)}(\undernu)}{\Psi^{(i)}(\undern)},
\end{split}
\ee
accumulates the Hxc potential energy contributions that are generated, for each cluster, by the interactions occurring beyond the impurity. 
\rev{The physical meaning of the density bi-functionals introduced in Eqs.~(\ref{eq:def_clusterization_kinetic_corr_ener_bifunc}) and (\ref{eq:beyond_imp_within_cluster_Hxc_pot_ener_bi_func}) is discussed in further detail in Appendix~\ref{appendix:clusterization_kin_corr_energy}.}\\

Finally, we stress that Eq.~(\ref{eq:exact_imp_chem_pots_relations}) is {\it invariant} under constant shifts in the density-functional Hxc potential, as expected. This is a consequence of the normalization of each bath orbital:
\be
\sum_k\inner{b^{(i)}(\undern)}{\chi_k}^2=\inner{b^{(i)}(\undern)}{b^{(i)}(\undern)}=1,\;\forall i.
\ee

\subsection{Practical local potential functional embedding theory}\label{sec:LPFET_from_exact_dens_func_embedding_theory}

The strongly local correlation regime is the one in which quantum embedding theory is expected to be accurate, for interacting electrons. In this extreme situation, we can proceed with the following density-functional approximations, 
\be
\mathcal{U}^{(i)}_{\rm Hxc}(\undernu,\undern)\approx 0,\;\;\forall i, 
\ee
and (see Eq.~(\ref{eq:clusterized_HK_func_exp})),
\be
F(\undern)\approx \sum_i\innerop{\Psi^{(i)}(\undern)}{\hat{t}^{(i)} + \hat{U}^{(i)}}{\Psi^{(i)}(\undern)},
\ee
meaning that $\bar{E}_{\rm Hxc}(\undern)\approx 0$. In addition, we can neglect kinetic contributions (because the local interactions dominate in the Hamiltonian), thus leading to 
\be
\tau^{(i)}_{\rm c}(\undernu,\undern)\approx 0,\;\;\forall i.
\ee
Consequently, under the above assumptions, the exact relation of Eq.~(\ref{eq:exact_imp_chem_pots_relations}), taken at the physical density $\undern=\undern^{\Psi_0}$, can be simplified as follows 
\be\label{eq:approx_relation_imp_chem_pots}
-\mu^{(j)}_{\rm imp}\underset{j\neq r}{\approx}-\mu^{(r)}_{\rm imp}
+\sum_k\left(\inner{b^{(r)}}{\chi_k}^2-\inner{b^{(j)}}{\chi_k}^2\right)v_k^{\rm Hxc}.\nonumber \\
\ee
If we (arbitrarily) adjust the Hxc potential such that
\be
\sum_k\inner{b^{(r)}}{\chi_k}^2v_k^{\rm Hxc}:=\mu^{(r)}_{\rm imp},
\ee
it then becomes uniquely defined (not up to a constant shift anymore) and, most importantly, {\it all} impurities fulfill the exact same relation, as a consequence of Eq.~(\ref{eq:approx_relation_imp_chem_pots}): 
\be\label{eq:final_LPFET_exp_chem_pots_wrt_Hxc_pot}
\mu^{(i)}_{\rm imp}\approx \sum_k\inner{b^{(i)}}{\chi_k}^2v_k^{\rm Hxc},\;\;\forall i.
\ee
Eq.~(\ref{eq:final_LPFET_exp_chem_pots_wrt_Hxc_pot}), where impurity chemical potentials are (approximately) expressed in terms of the Hxc potential, is the second key result of the present work. Combined with Eqs.~(\ref{eq:dens_func_cluster_Hamil_main_text}) and (\ref{eq:exact_dens_constraint_cluster}), it lays the foundation of the LPFET approach, where the basic variable is the (local) Hxc potential ${\boldv}^{\rm Hxc}=\left\{v_k^{\rm Hxc}\right\}$. Indeed, for a trial ${\boldv}^{\rm Hxc}$, the embedding cluster wavefunctions are determined as follows, 
\be\label{eq:LPFET_cluster_wf_variational_calc}
\Psi^{(i)}\left(\boldv^{\rm Hxc}\right)
=\argmin_{\Psi\rightarrow 2}\innerop{\Psi}{\hat{P}^{(i)}\hat{H}\hat{P}^{(i)}-\mu_{\rm LPFET}^{(i)}\hat{n}_i}{\Psi}, \nonumber \\
\ee
where the impurity-{\it dependent} LPFET chemical potential reads 
\be\label{eq:ind_LPFET_chem_pot} 
\mu_{\rm LPFET}^{(i)}\equiv \mu_{\rm LPFET}^{(i)}\left(\boldv^{\rm Hxc}\right)  =\sum_{k}\inner{b^{(i)}}{\chi_k}^2v^{\rm Hxc}_k.
\ee
Let us stress that, like the full-size KS wavefunction,
\be\label{eq:find_full-size_KS_det_from_v_Hxc}
\Phi^{\rm KS}\left(\boldv^{\rm Hxc}\right)
=\argmin_{\Phi\rightarrow N}\innerop{\Phi}{
\hat{T}+\hat{V}^{\rm ext}+\sum_iv_i^{\rm Hxc}\hat{n}_i}{\Phi}, \nonumber \\
\ee
the projector $\hat{P}^{(i)}\equiv \hat{P}^{(i)}\left(\boldv^{\rm Hxc}\right)$ in Eq.~(\ref{eq:LPFET_cluster_wf_variational_calc}) and the bath orbital $b^{(i)}\equiv b^{(i)}\left(\boldv^{\rm Hxc}\right)$ in Eq.~(\ref{eq:ind_LPFET_chem_pot}) are functionals of ${\boldv}^{\rm Hxc}$. However, unlike $\mu_{\rm LPFET}^{(i)}$, they are both invariant under constant shifts in ${\boldv}^{\rm Hxc}$. The overall embedding procedure is finally made self-consistent through the density constraints of Eqs.~(\ref{eq:local_dens_mapping_KS}) and (\ref{eq:exact_dens_constraint_cluster}):    
\be\label{eq:LPFET_dens_const_sc_loop}
\myexpval{\hat{n}_i}{\Phi^{\rm KS}\left(\boldv^{\rm Hxc}\right)}
\underset{\rm LPFET}{\overset{!}{=}}\myexpval{\hat{n}_i}{\Psi^{(i)}\left(\boldv^{\rm Hxc}\right)},\;\;\forall i.
\ee
Note that Eq.~(\ref{eq:LPFET_dens_const_sc_loop}) ensures that the correct total number $N$ of electrons in the full system is recovered from all the embedding clusters. In the special case of a uniform density profile ($v_k^{\rm Hxc}=v^{\rm Hxc}$), we recover the result of Ref.~\citenum{sekaran2022local}, namely $\mu_{\rm LPFET}^{(i)}=v^{\rm Hxc},\;\forall i$. 
\\

Comparison with DMET should obviously be made at this point. For that purpose, we can use the value of the Hxc potential on some reference site (the residual site introduced previously, for example) as a global chemical potential value $v^{\rm Hxc}_r=:\mu_{\rm glob}$ that has to be determined.
Deviations $\Delta{\boldv}^{\rm Hxc}\equiv\left\{\Delta v^{\rm Hxc}_k=v^{\rm Hxc}_k-v^{\rm Hxc}_r\right\}$ from that reference value will completely define ${\boldv}^{\rm Hxc}$. By rewriting Eq.~(\ref{eq:ind_LPFET_chem_pot}) as follows,  
\begin{eqnarray}
\mu_{\rm LPFET}^{(i)}&=&\sum_{k}\inner{b^{(i)}}{\chi_k}^2\left(\Delta v^{\rm Hxc}_k+\mu_{\rm glob}\right)\nonumber
\\
&=&\sum_{k}\inner{b^{(i)}}{\chi_k}^2\Delta v^{\rm Hxc}_k+\left(\sum_{k}\inner{b^{(i)}}{\chi_k}^2\right)\mu_{\rm glob} \nonumber
\\
&=&\sum_{k}\inner{b^{(i)}}{\chi_k}^2\Delta v^{\rm Hxc}_k+\inner{b^{(i)}}{b^{(i)}}\mu_{\rm glob},
\end{eqnarray}
we deduce from the normalization of the bath orbital (see Eq.~(\ref{eq:bath_from_1RDM})) the expression below,
\be\label{eq:rewriting_LPFET_imp_chem_pot_delta_plus_global}
\mu_{\rm LPFET}^{(i)}=\sum_{k}\inner{b^{(i)}}{\chi_k}^2\Delta v^{\rm Hxc}_k+\mu_{\rm glob},
\ee
where $b^{(i)} \equiv b^{(i)}\left(\Delta{\boldv}^{\rm Hxc}\right)$. From the above equation we can immediately identify a major difference between LPFET and DMET. Indeed, in the latter case, the exact same chemical potential value $\mu_{\rm glob}$ is used in {\it all} clusters~\cite{wouters2016practical},
\be\label{eq:DMET_emb_cluster_wf_opti}
\begin{split}
&\Psi^{(i)}\left(\boldv^{\rm Hxc}\right)
{\rightarrow} \Psi^{(i)}\left(\Delta\boldv^{\rm Hxc},\mu_{\rm glob}\right)
\\
&\underset{\rm DMET}
{=}
\argmin_{\Psi\rightarrow 2}\innerop{\Psi}{\hat{P}^{(i)}\hat{H}\hat{P}^{(i)}-\mu_{\rm glob}\hat{n}_i}{\Psi}.
\end{split}
\ee
By imposing, like in LPFET, a local density mapping constraint, DMET becomes strictly speaking a DET in which the self-consistency loop reads as follows,
\be
\myexpval{\hat{n}_i}{\Phi^{\rm KS}\left(\Delta\boldv^{\rm Hxc}\right)}
\underset{\rm DET}{\overset{!}{=}}\myexpval{\hat{n}_i}{\Psi^{(i)}\left(\Delta\boldv^{\rm Hxc},\mu_{\rm glob}\right)},\;\;\forall i.
\ee
In conclusion, LPFET and DET differ in the impurity chemical potential expression. The former ($\mu_{\rm LPFET}^{(i)}$) depends on the embedded orbital (through the associated bath orbital) while the other ($\mu_{\rm DET}=\mu_{\rm glob}$) does not: 
\be
\mu_{\rm LPFET}^{(i)}-\mu_{\rm DET}=\sum_{k}\inner{b^{(i)}}{\chi_k}^2\Delta v^{\rm Hxc}_k.
\ee
Let us stress that, in the limit of non-interacting electrons (${\boldv^{\rm Hxc}}=\Delta{\boldv^{\rm Hxc}}\equiv 0$), DET and LPFET are identical and exact.\\  

For completeness, we should finally mention that in both LPFET and DET approaches, unlike in SDE~\cite{mordovina2019self}, no KS inversion is performed within the embedding clusters. Instead, the Hxc potential values will all be updated at once in the full-size KS system, at each iteration (see Sec.~\ref{sec:comput_details} for further practical details).

\subsection{Interpretation of LPFET and alternative derivation}\label{sec:rationale_and_extensions_LPFET}

As shown in Appendix~\ref{appendix:relations_chem_pots_partial-int_bath} and summarized in Eq.~(\ref{eqapp:general-PI_bath_LPFET}), the LPFET impurity chemical potential expression of Eq.~(\ref{eq:ind_LPFET_chem_pot}), which in fact only involves the Hxc potential values on the environment of the impurity (since the impurity orbital $\chi_i$ is orthogonal to the bath orbital $b^{(i)}$, by construction), \ie,
\be\label{eq:simplified_imp_chem_pot_LPFET}
\mu_{\rm LPFET}^{(i)}=\sum_{k}\inner{b^{(i)}}{\chi_k}^2v^{\rm Hxc}_k=\sum_{k\neq i}\inner{b^{(i)}}{\chi_k}^2v^{\rm Hxc}_k, \nonumber \\
\ee
holds whether the bath is non-interacting (NIB), partially-interacting, or fully-interacting, like in Eq.~(\ref{eq:LPFET_cluster_wf_variational_calc}). At first sight, this conclusion will probably look surprizing to DMET practitioners since standard NIB and IB formulations of DMET differ precisely on this point~\cite{wouters2016practical}. The underlying reason, formulated in Ref.~\citenum{wouters2016practical}, for using a global (\ie, impurity-independent) chemical potential value in the IB case, in place of the impurity-dependent expression of Eq.~(\ref{eq:simplified_imp_chem_pot_LPFET}), is the fact that double counting of electron correlation should be prevented. The derivation of Appendix~\ref{appendix:relations_chem_pots_partial-int_bath}, which relies on an in-principle exact (and therefore double-counting-free) density-functional embedding theory in which well-identified density-functional approximations are made, all based on the assumption that electron correlation is strong and local [see Eqs.~(\ref{eq:partially_int_dens_fun_Hamil}) and (\ref{eqapp:partial_electronic_rep}), and the comment that follows, for the exact theory, and Eq.~(\ref{eqapp:part-int-pot-corr_approx}), for the local density-functional approximation that leads to LPFET], offers a different viewpoint on the question. First of all, it confirms that the expression of Eq.~(\ref{eq:simplified_imp_chem_pot_LPFET}), which is used in the NIB formulation of DMET (see Eqs.~(19) and (20) of Ref.~\citenum{wouters2016practical}), gives a sound approximation to the impurity chemical potential. In this case, it is simply deduced from the projection of the full-size interacting-fragment-only density-functional Hamiltonian (see Eqs.~(\ref{eq:partially_int_dens_fun_Hamil}), (\ref{eqapp:range_int_NIB_case}), (\ref{eq:app_NIB_partial_interaction}), and (\ref{eqapp:part-int-pot-corr_approx})) onto the embedding cluster. Secondly, assuming that electron correlation is essentially local, Appendix~\ref{appendix:relations_chem_pots_partial-int_bath} shows that the expression of Eq.~(\ref{eq:simplified_imp_chem_pot_LPFET}) still ensures that the embedded impurity has the correct density even when interactions in the environment of the fragment are taken into account (thus making the bath interacting), as depicted in a more formal way in Eqs.~(\ref{eqapp:approx_relation_imp_chem_pots-partIB})--(\ref{eqapp:general-PI_bath_LPFET}).\\    

\rev{Further physical insight can be obtained from an alternative (but equivalent) approach whose detailed derivation is provided in Appendix~\ref{app:alternative_deriv_LPFET}. Based on the approximate evaluation of the Hxc energy as a {\it local density} functional, it highlights the underlying approximations of LPFET and definitely removes doubts about potential double counting issues. This more intuitive formulation is also expected to be useful when extending the embedding to multiple-orbital fragments~\cite{yalouz2022quantum,sekaran2023unified} and/or ensembles of ground and excited states\cite{Cernatic_2021}, for example. 
}

\section{Implementation and computational details}\label{sec:comput_details}

In this work, we apply the LPFET approach (its IB formulation, as depicted in Eq.~(\ref{eq:LPFET_cluster_wf_variational_calc})) to two simple but nontrivial illustrative systems for which exact (FCI) solutions can be computed, for comparison. The first is a non-uniform finite Hubbard ring consisting of six lattice sites ($L = 6$) at half-filling, with periodic boundary conditions and a nearest-neighbor hopping parameter set to $t = 1$. The corresponding Hamiltonian reads
\be
\begin{split}
\hat{H}&=-t\sum^{L-1}_{i=0}\sum_{\sigma}\left(\hat{c}_{i\sigma}^\dagger\hat{c}_{i+1\sigma}+\hat{c}_{i+1\sigma}^\dagger\hat{c}_{i\sigma}\right)
\\
&\quad
+U\sum^{L-1}_{i=0}\hat{n}_{i\uparrow}\hat{n}_{i\downarrow}
+\sum^{L-1}_{i=0}v^{\rm ext}_i\hat{n}_i
,
\end{split}
\ee
where $\left\{v^{\rm ext}_i\right\}_{0\leq i\leq 5}\equiv \left\{-1, 2, -2, 3, -3, 1\right\}$. While the latter (fixed) external potential ensures that each fragment orbital ``sees'' a different environment, which is optimal for testing the robustness of the LPFET approach, varying the on-site interaction/hopping ($U/t$) ratio allows to control the strength of electron correlation in the system. The second system is an {\it ab initio} hydrogen chain, composed of six equispaced hydrogen atoms aligned linearly and described using a minimal STO-3G atomic orbital basis set. Various bond distances are considered to probe a wide range of correlation regimes. Symmetrically orthogonalized atomic orbitals (OAOs) were constructed following Löwdin’s method~\cite{lowdin1950non} and used as localized orbital basis in quantum embedding calculations.\\

The LPFET algorithm, whose theoretical foundation has been discussed in detail in Secs.~\ref{sec:LPFET_from_exact_dens_func_embedding_theory} and \ref{sec:rationale_and_extensions_LPFET}, can be installed and executed following the LPFET github repository~\cite{lpfet_github} \rev{(commit \texttt{306f90e})}. It can be summarized as follows.

\begin{enumerate}
    \item  \textbf{Initialization and full-size KS calculation}: Even though a more educated choice would be desirable (to speed up convergence), we simply initialize the Hxc potential to zero (${\boldv^{\rm Hxc}}\equiv 0$) for convenience but also for convergence analysis. We then diagonalize the one-electron KS Hamiltonian [see Eq.~(\ref{eq:find_full-size_KS_det_from_v_Hxc})] to obtain the (delocalized) molecular KS orbitals and energies, and the resulting ground-state density matrix (from the Aufbau principle). Unlike in LPFET of uniform Hubbard lattices in the thermodynamic limit (for which only the physical chemical potential value was given~\cite{sekaran2022local}), the total number of electrons in the system is supposed to be known (and fixed to an integer) in the present context.  

    \item \textbf{KS Density Matrix and Embedding Clusters Construction}: We write the KS density matrix in the lattice (or localized orbital) representation, which gives access to the KS density profile $\left\{\myexpval{\hat{n}_i}{\Phi^{\rm KS}\left(\boldv^{\rm Hxc}\right)}\right\}$, and determine, for each single-orbital fragment, individually, the corresponding embedding cluster Hamiltonian [see Eq.~(\ref{eq:LPFET_cluster_wf_variational_calc}) and Eq.~(\ref{eq:ind_LPFET_chem_pot}), where the LPFET impurity chemical potential is expressed explicitly in terms of the Hxc potential]. A Householder transformation~\cite{sekaran2021householder,yalouz2022quantum,sekaran2023unified} is used for that purpose.

    \item \textbf{Solutions to each Embedding Cluster}: 
    The (two-electron in the present case) ground-state Schr\"{o}dinger equation is solved for each embedding cluster, thus giving access to the correlated occupation $\myexpval{\hat{n}_i}{\Psi^{(i)}\left(\boldv^{\rm Hxc}\right)}$ of each embedded impurity.

    \item \textbf{Convergence Check}: 
    The occupations of each embedded impurity $\left\{\myexpval{\hat{n}_i}{\Psi^{(i)}\left(\boldv^{\rm Hxc}\right)}\right\}$ are compared with those of the KS system $\left\{\myexpval{\hat{n}_i}{\Phi^{\rm KS}\left(\boldv^{\rm Hxc}\right)}\right\}$. If they do not match, the Hxc potential is updated using a cost function implemented in Python, and we return to the first step above. The cost function quantifies the discrepancy between the KS and embedded impurity densities. At convergence, the two density profiles should match, according to Eq.~(\ref{eq:LPFET_dens_const_sc_loop}), thus ensuring that, ultimately, the combined embedding clusters reproduce the correct (integer) number of electrons in the system.
    \end{enumerate}
\rev{The corresponding LPFET algorithm chart reads as follows. 
\begin{algorithm}[H]
\caption{LPFET Self-Consistent Embedding Loop}
\begin{algorithmic}[1]
\State \textbf{Input:} Hamiltonian $H$, number of electrons $N_{\rm el}$
\State \textbf{Output:} Converged $\boldv^{\rm Hxc}$, cluster densities

\State $\boldv^{\rm Hxc} \gets 0$ \Comment{Initialize Hxc potential}
\State converged $\gets$ False

\While{not converged}
    \State \textbf{Step 1: Initialization and Full-Size KS Calculation}
        \State Diagonalize $H_{\rm KS}(\boldv^{\rm Hxc})$ to obtain $\Phi^{\rm KS}$, orbital energies $\epsilon_i^{\rm KS}$, $\gamma^{\rm KS}$
    
    \State \textbf{Step 2: KS Density Matrix and Embedding Clusters Construction}
        \For{each site $i$}
            \State $n_i^{\rm KS} \gets$ local density from $\gamma^{\rm KS}$
            \State Apply Householder transformation and build $H_{\rm cluster}^{(i)}$
        \EndFor

    \State \textbf{Step 3: Solve Embedding Clusters}
        \For{each cluster $i$}
            \State Solve the 2-electron Schr\"odinger equation
            \State $n_i^{\rm cluster} \gets \langle \Psi^{(i)} (\mathbf{v}^{\rm Hxc})| \hat{n}_i | \Psi^{(i)}(\boldv^{\rm Hxc}) \rangle$
        \EndFor

    \State \textbf{Step 4: Convergence Check}
        \If{ $|n_i^{\rm cluster} -n_i^{\rm KS}|^2 \leq \epsilon$ with $\epsilon=10^{-9}$}
            \State converged $\gets$ True
        \Else
            \State Update $\boldv^{\rm Hxc}$ using a classical optimizer (L-BFGS-B)
        \EndIf
\EndWhile

\State \textbf{Return:} $\boldv^{\rm Hxc}$, cluster densities
\end{algorithmic}
\end{algorithm}

}

\section{Results and discussion} \label{Results}
\subsection{Non-uniform Hubbard ring}\label{Hubbard_ring}

\begin{figure}
    \centering
    \includegraphics[width=1.17\linewidth]{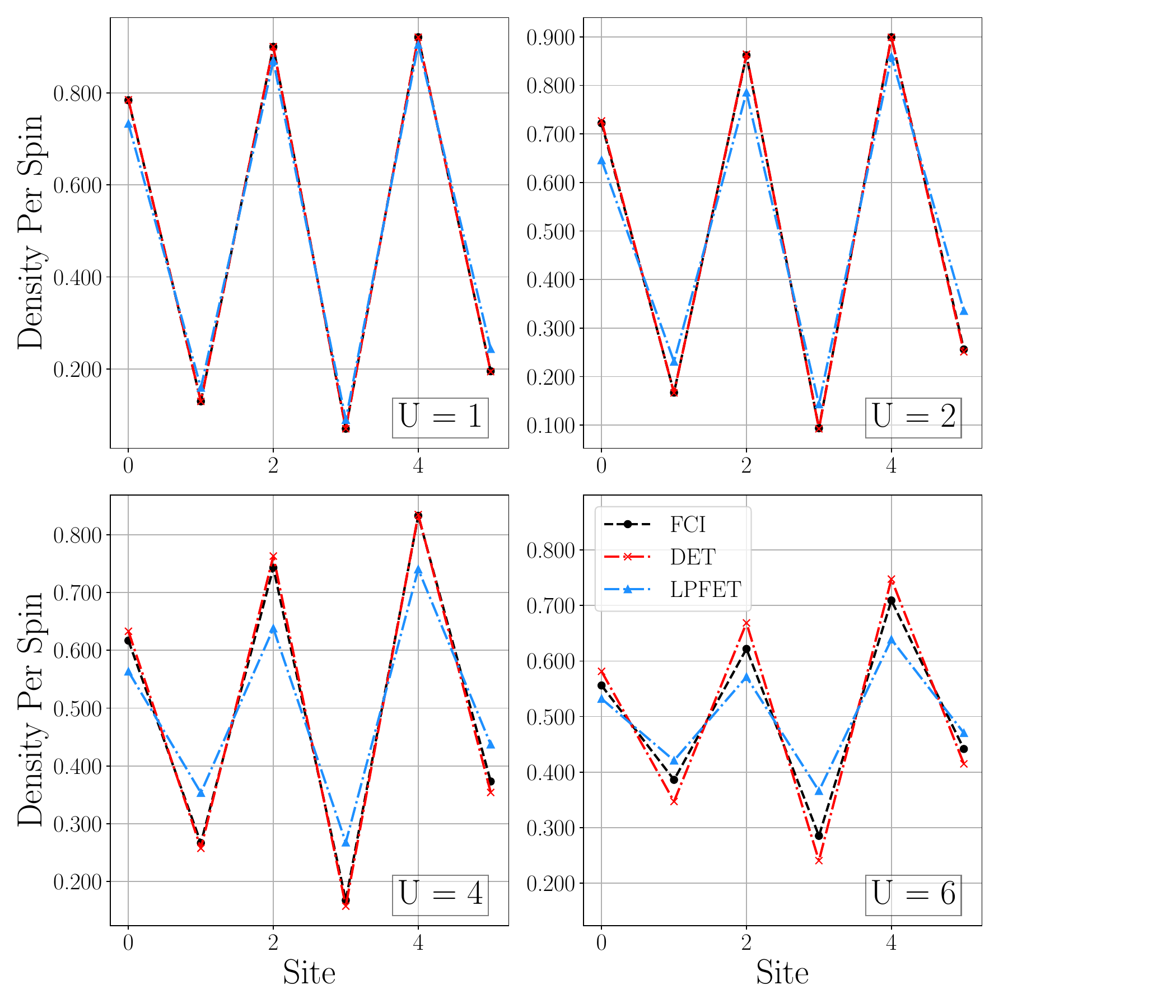}
        \caption{Local density profiles obtained from self-consistent LPFET and DET calculations for weakly and moderately correlated half-filled 6-site Hubbard rings. The hopping parameter is set to $t=1$ and the external potential is fixed with $\max_i \left\{\vert v_i^{\rm ext}\vert\right\}=3$. Comparison is made with FCI. See text for further details.}
    \label{fig:weakly_strongly_correlated}
\end{figure}

\begin{figure}
    \centering
    \includegraphics[width=1.175\linewidth]{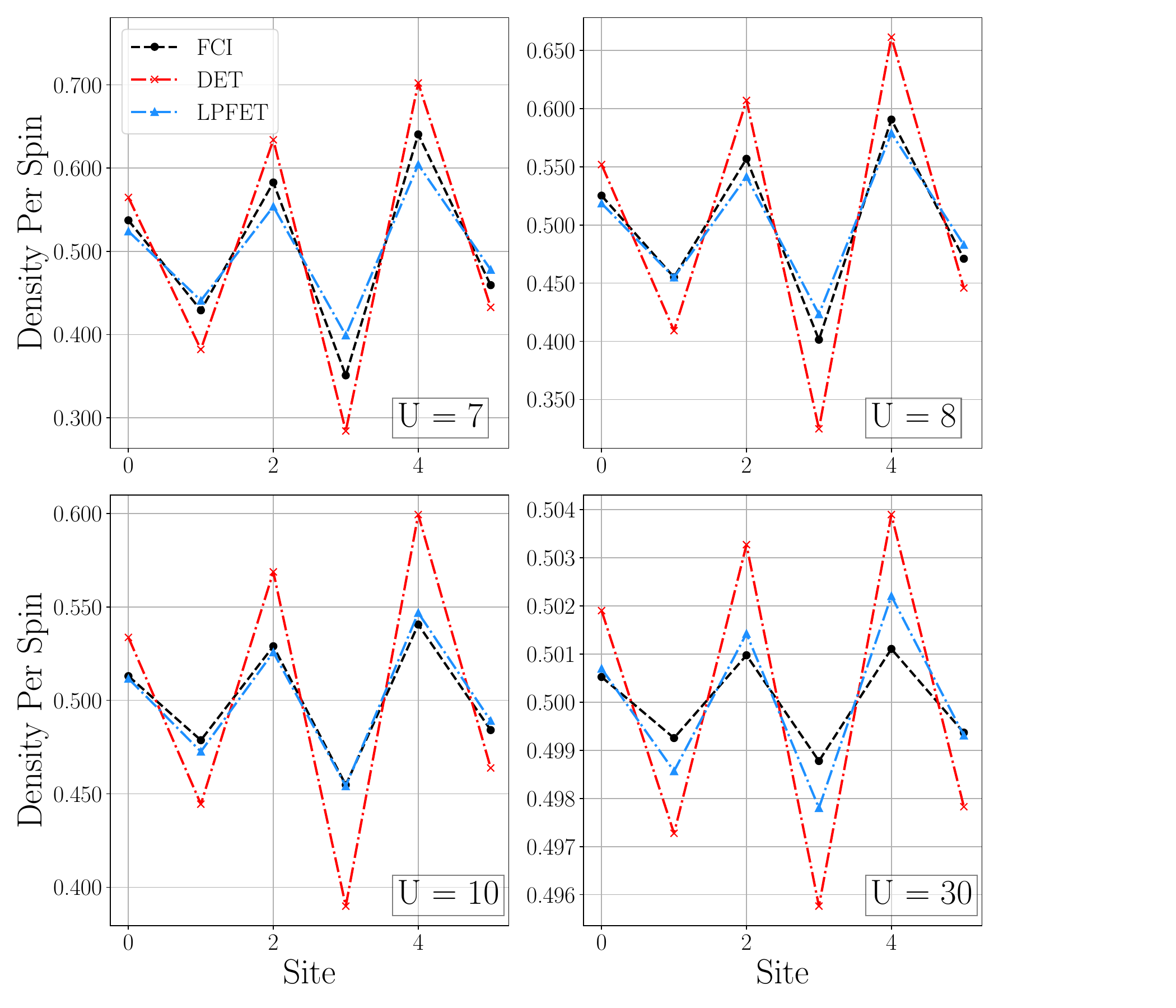}
    \caption{Same as Fig.~\ref{fig:weakly_strongly_correlated} for stronger correlation regimes.}
    \label{intermedite_correlated}
\end{figure}

\begin{figure}
    \centering
    \includegraphics[width=1.17\linewidth]{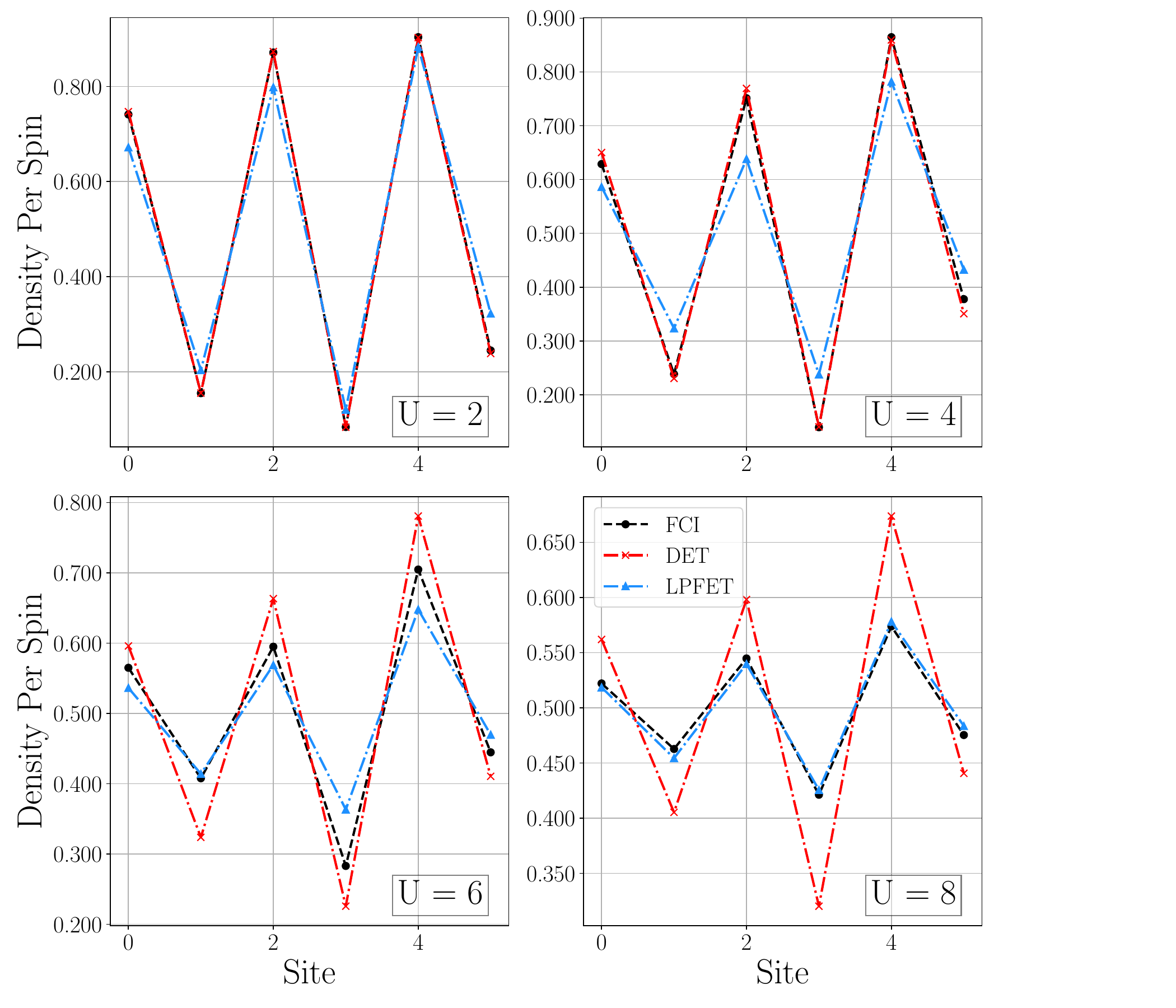}
        \caption{Local density profiles obtained from self-consistent LPFET and DET calculations for weakly and moderately correlated half-filled 6-site Hubbard ladders. Comparison is made with FCI. See text for further details.}
    \label{fig:weakly_strongly_correlated_2D}
\end{figure}

We begin our analysis with the non-uniform six-site Hubbard ring at half-filling. The performance of LPFET is assessed by comparison with FCI, which provides an exact solution to the model, and DET. Density profiles are shown in Figs.~\ref{fig:weakly_strongly_correlated} and \ref{intermedite_correlated} for various on-site interaction strengths. As expected, LPFET and DET give similar results close to the non-interacting limit (as seen for $U/t=1$, for example) but, as electron correlation becomes stronger (up to $U/t=4$), substantial differences between the LPFET and FCI density profiles can be seen (errors are not dramatic though) while DET remains very accurate. Interestingly, when entering stronger correlation regimes (from $U/t=6$), where LPFET finds its theoretical justification, this tendency reverses. In particular, LPFET outperforms DET (whose density deviates by 0.05 from the FCI) when $U/t=8$ or $U/t=10$. For even stronger interaction strengths ($U/t=30$, for example), LPFET is still more accurate than DET but the error in density (of the order of 0.003) is reduced. 
\rev{
For completeness, we show in Fig.~\ref{fig:weakly_strongly_correlated_2D} the density profiles obtained in various correlation regimes for a two-dimensional (2D) version of our non-uniform (six-site and half-filled) Hubbard model. Starting from the one-dimensional (1D) ring model, we simply introduced the exact same hopping term $t=1$ between sites 1 and 4, thus giving a ladder version of the model. No major difference can be seen between the 1D and 2D models in this context.   
}
The numerical results compiled in Figs.~\ref{fig:weakly_strongly_correlated},~\ref{intermedite_correlated}, and \ref{fig:weakly_strongly_correlated_2D} nicely confirm that, when electron correlation is strong and local, each impurity chemical potential can be expressed explicitly in terms of the (local) Hxc potential, a property that is exploited in LPFET. Its impact on the self-consistent evaluation of the Hxc potential will be discussed further in the following. 
\begin{figure}
    \includegraphics[width=1.18\linewidth]{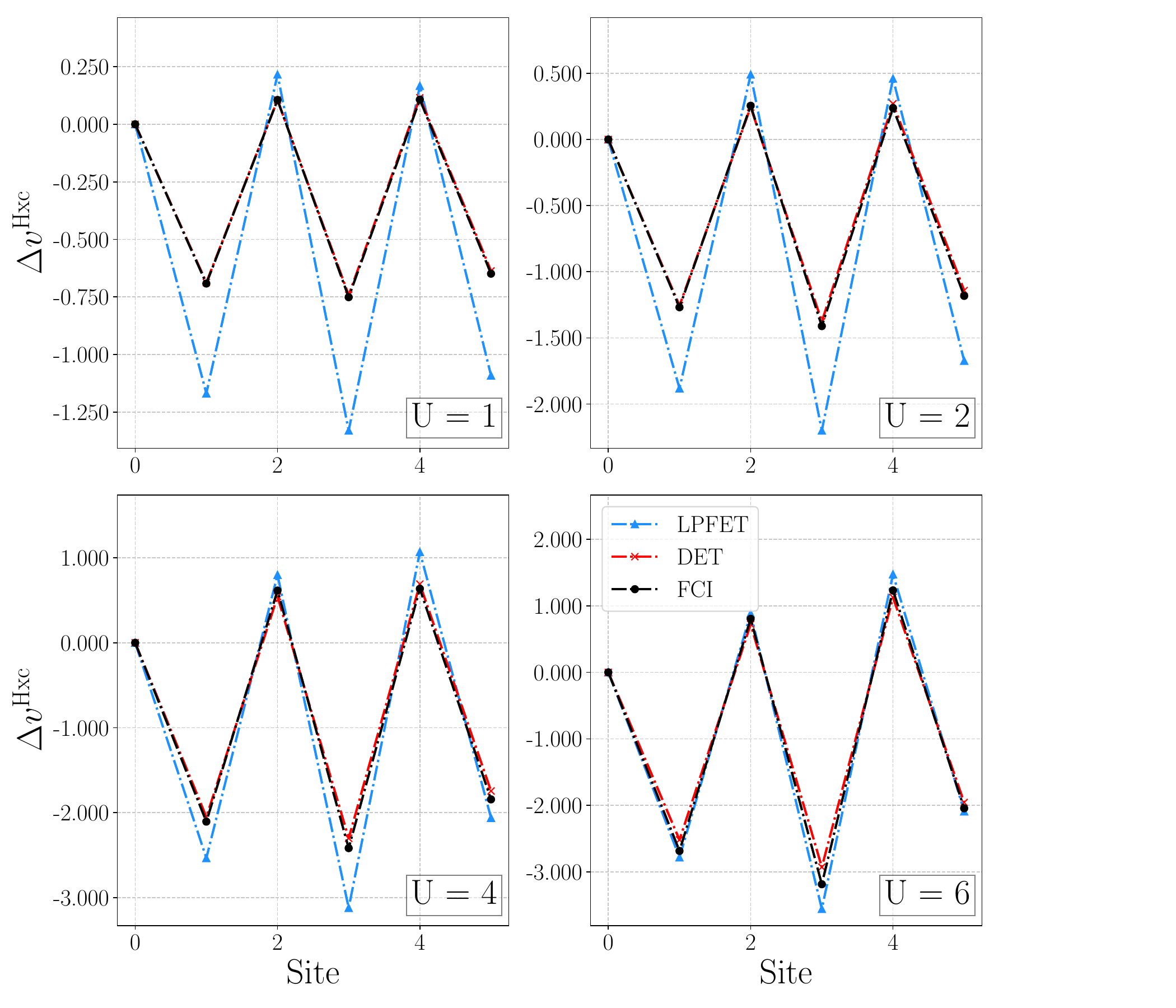}
    \caption{Self-consistently converged Hxc potential profiles relative to site 0 ($\Delta v^{\text{Hxc}}_i=v^{\text{Hxc}}_i-v^{\text{Hxc}}_0$) obtained for the 6-site Hubbard ring with LPFET and DET methods in weakly and moderately correlated regimes. Comparison is made with the exact Hxc potential that has been determined from the FCI density profile.} 
    \label{V_hxc_potentiel}
\end{figure}
\begin{figure}
    \centering
    \includegraphics[width=1.75\linewidth]{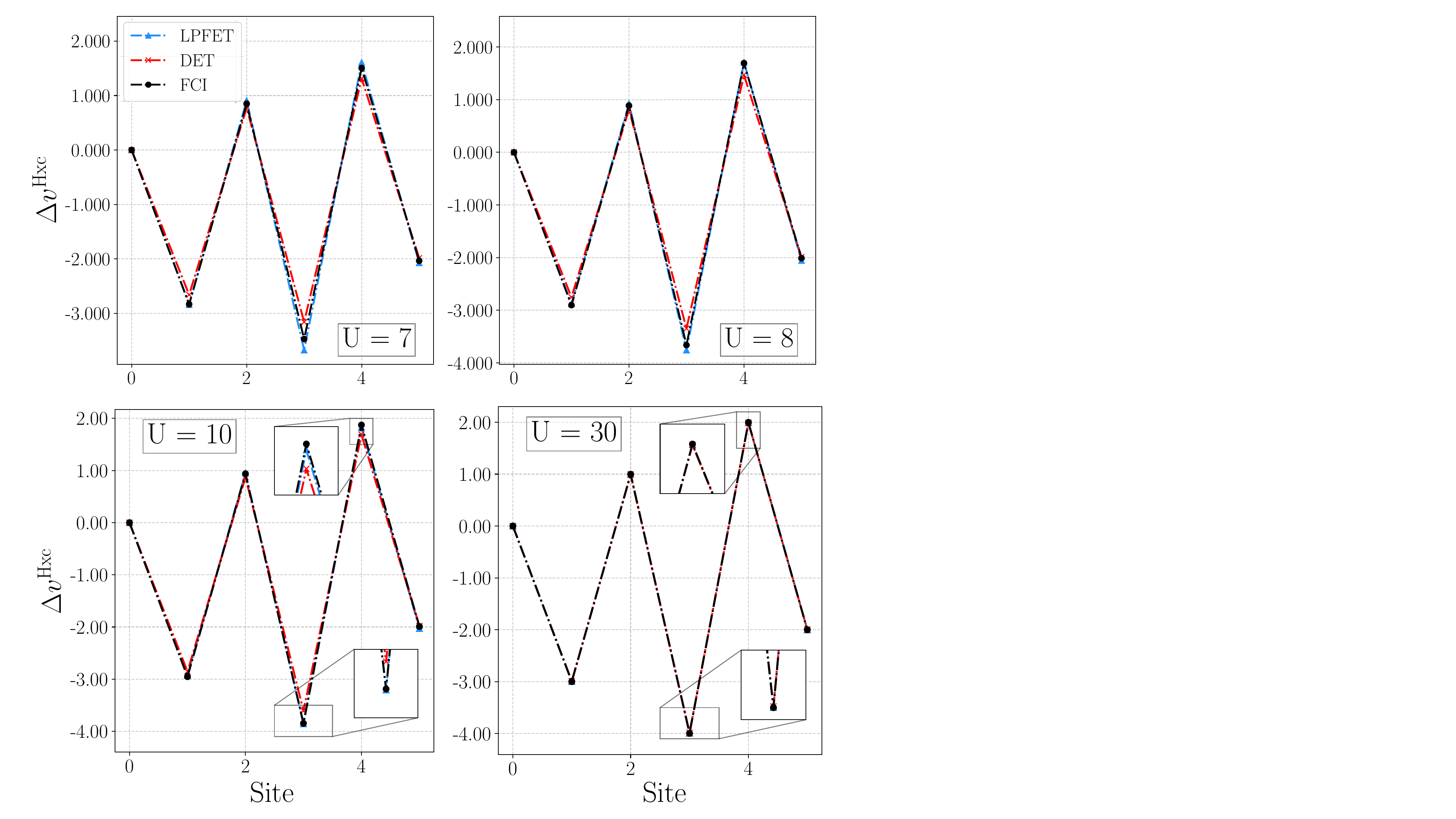}
    \caption{Same as Fig.~\ref{V_hxc_potentiel} for stronger correlation regimes.}
    \label{V_hxc_potentiel_intermediate}
\end{figure}
Finally, note that these observations are consistent with the converged Hxc potential profiles shown in Figs.~\ref{V_hxc_potentiel} and~\ref{V_hxc_potentiel_intermediate}. For example, we clearly see that the LPFET Hxc potential is increasingly overattractive on sites 1, 3, and 5, as the interaction strength increases up to $U/t=4$, thus leading to too high densities (see Fig.~\ref{fig:weakly_strongly_correlated}). In the transition $6\leq U/t\leq 8$ regime, differences between LPFET, DET and FCI Hxc potentials are much less visible. It still appears, on site 3 (4), for example, that the DET Hxc potential is now under(over)attractive, as expected from the above analysis of the density profiles. This feature persists when $U/t=10$, as shown in the insets of Fig.~\ref{V_hxc_potentiel_intermediate}. As electron correlation increases further, the LPFET, DET and FCI Hxc potentials become almost indistinguishable.\\


\begin{figure}
    \centering
    \includegraphics[width=1\linewidth]{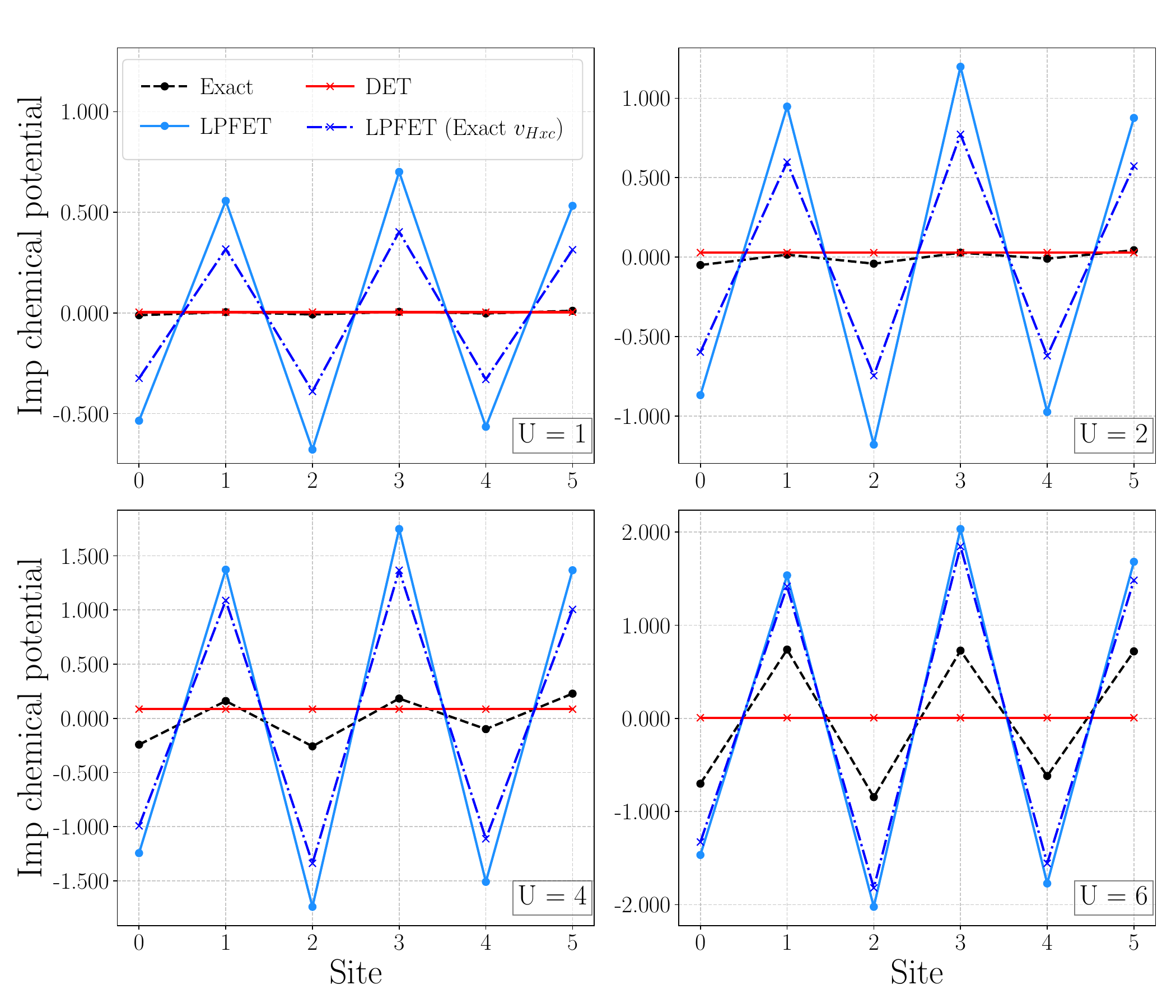}
    \caption{Self-consistently converged LPFET impurity chemical potentials computed for each embedded impurity of the 6-site Hubbard ring in weakly and moderately correlated regimes. Comparison is made with the exact (FCI) and global (self-consistently converged) DET values. For analysis purposes, the impurity chemical potential values directly computed from the exact (FCI) Hxc potential, according to Eq.~(\ref{eq:rewriting_LPFET_imp_chem_pot_delta_plus_global}), are also plotted. See text for further details.}
    \label{fig:mu_comparison}
\end{figure}


\begin{figure}
    \centering
    \includegraphics[width=1\linewidth]{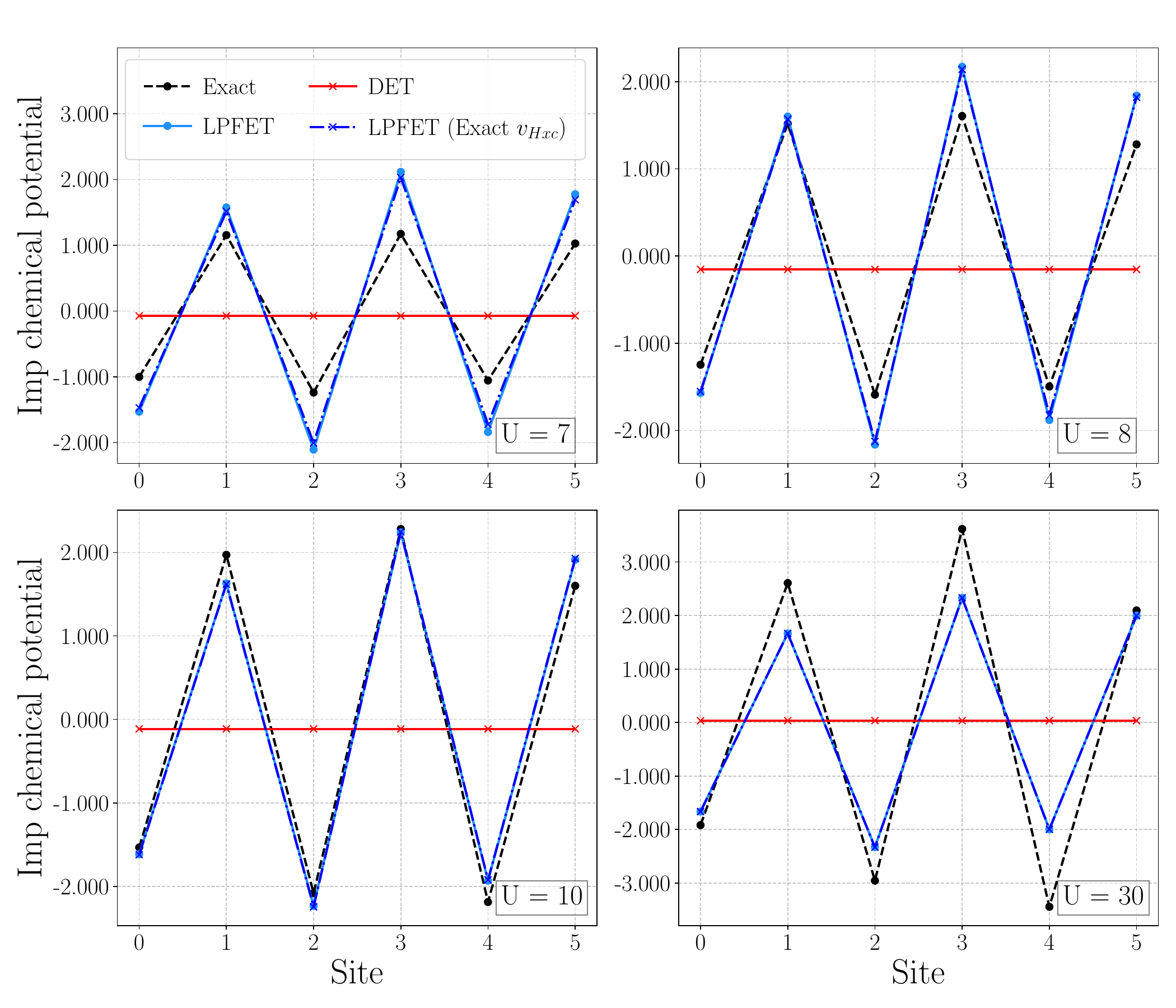}
    \caption{Same as Fig.~\ref{fig:mu_comparison} for stronger correlation regimes.}
    \label{mu_comparison_interdediate}
\end{figure}
In order to further analyze the performance of LPFET and DET methods, we show in Figs.~\ref{fig:mu_comparison} and~\ref{mu_comparison_interdediate} impurity chemical potential values obtained for each embedded impurity at various levels of approximation. Comparison is made with the self-consistently converged global (impurity-independent) chemical potential value of DET. 
\rev{
In this simple model, exact impurity chemical potentials can be evaluated from the reference FCI calculation. Indeed, from the FCI density profile, we can determine the exact Hxc potential through reverse engineering (a cost function is used for that purpose). As a result, we can determine the exact KS density matrix, thus providing the exact bath for each embedded impurity, according to Eq.~(\ref{eq:bath_from_1RDM}). The exact impurity chemical potential is finally determined such that the local embedded impurity density matches that of the KS system or, equivalently, the FCI density. For analysis purposes, we also evaluated the LPFET impurity chemical potential from the exact Hxc potential, according to Eq.~(\ref{eq:rewriting_LPFET_imp_chem_pot_delta_plus_global}). Note that, in this case, reverse engineering only provides Hxc potential differences (with respect to some reference site). We determined the remaining global shift such that the density profile obtained from the embedding clusters is as close as possible to the KS one (a cost function was used for that purpose).}
We first notice in Fig.~\ref{fig:mu_comparison} that, for the relatively small $U/t=1$ interaction strength, the exact impurity chemical potential profile is almost uniform and equal to zero, which is very well reproduced by DET. 
In this case, the one-electron mean-field (Hubbard) Hamiltonian, whose exchange potential is local in the lattice representation, like in the KS Hamiltonian, is expected to be a good approximation to the true interacting Hamiltonian. Consequently, the (IB in the present case) embedding is essentially exact, with no need for additional impurity chemical potential corrections (see the end of Sec.~\ref{sec:DMET_review}). 
On the other hand, LPFET introduces unnecessary impurity chemical potential corrections within each cluster, according to Eq.~(\ref{eq:simplified_imp_chem_pot_LPFET}), where the Hx potential is expected to be the leading contribution. 
\rev{The key problem here is that LPFET was derived as an approximation (to an exact theory that involves the Hartree, exchange, and correlation contributions to the KS potential as a whole, ${\bf v}^{\rm Hxc}$) that is expected to be justified only in the strongly correlated regime. Applying this approximation to weakly correlated regimes essentially consists in neglecting the correlation potential, which is obviously highly questionable from a purely theoretical point of view.}
In practice, this makes the embedded impurity sites over (or under) attractive, thus leading to the deviations in density profile of LPFET from FCI or DET (see Fig.~\ref{fig:weakly_strongly_correlated}). Interestingly, when the exact Hxc potential is used in place of the converged LPFET one, according to Eq.~(\ref{eq:rewriting_LPFET_imp_chem_pot_delta_plus_global}), the impurity chemical potential profile is (not completely but) more uniform. This tends to indicate that self-consistency deteriorates the LPFET densities of weakly correlated systems. A perturbative analysis, in the spirit of Ref.~\citenum{cances2025analysis}, would probably give further insight into how to improve LPFET in the weakly correlated regime. 
\rev{A more pragmatic alternative, inspired by the use of the exact (\ie, Hartree-Fock-like) exchange energy expression in the context of (generalized) KS-DFT, would consist in combining the non-local (in the lattice) exact exchange potential with a local correlation potential. Deriving such a generalized LPFET is expected to relate the impurity chemical potential to ${\bf v}^{\rm c}$ only, thus hopefully fixing the problem for weak correlations (while preserving the nice features of LPFET in stronger correlation regimes).}
This is left for future work. Note that, while the errors of LPFET on the impurity chemical potential profile are enhanced when the interaction strength increases up to $U/t=4$, we also notice that the exact impurity chemical potential profile increasingly deviates from uniformity. 
This is a key observation that confirms numerically the need for a specific chemical potential for each embedded fragment orbital, individually. This feature, which is enhanced by increasing further the interaction strength (see Fig.~\ref{mu_comparison_interdediate}), is completely missed by DET. On the other hand, LPFET can capture this effect, by construction. 
It becomes quantitatively correct when electron correlation is strong, a regime in which it finds its theoretical justification. In this case, we note that using the exact or converged LPFET Hxc potential brings no difference, thus confirming the relevance of the LPFET impurity chemical potential ansatz. As a final comment, we should emphasize that the impurity chemical potential is used within the interacting embedding cluster. As a result, the deviations from uniformity, as exhibited in the $U/t=30$ panel of Fig.~\ref{mu_comparison_interdediate}, become less critical (because they are much smaller than the interaction strength) in the evaluation of the correlated embedded impurity occupation. This explains why, ultimately, LPFET and DET do not differ substantially anymore in the Hxc potential (and density) profile, as seen from the bottom right panels of Figs.~\ref{intermedite_correlated} and \ref{V_hxc_potentiel_intermediate}. 

\begin{figure}
    \includegraphics[width=1\linewidth]{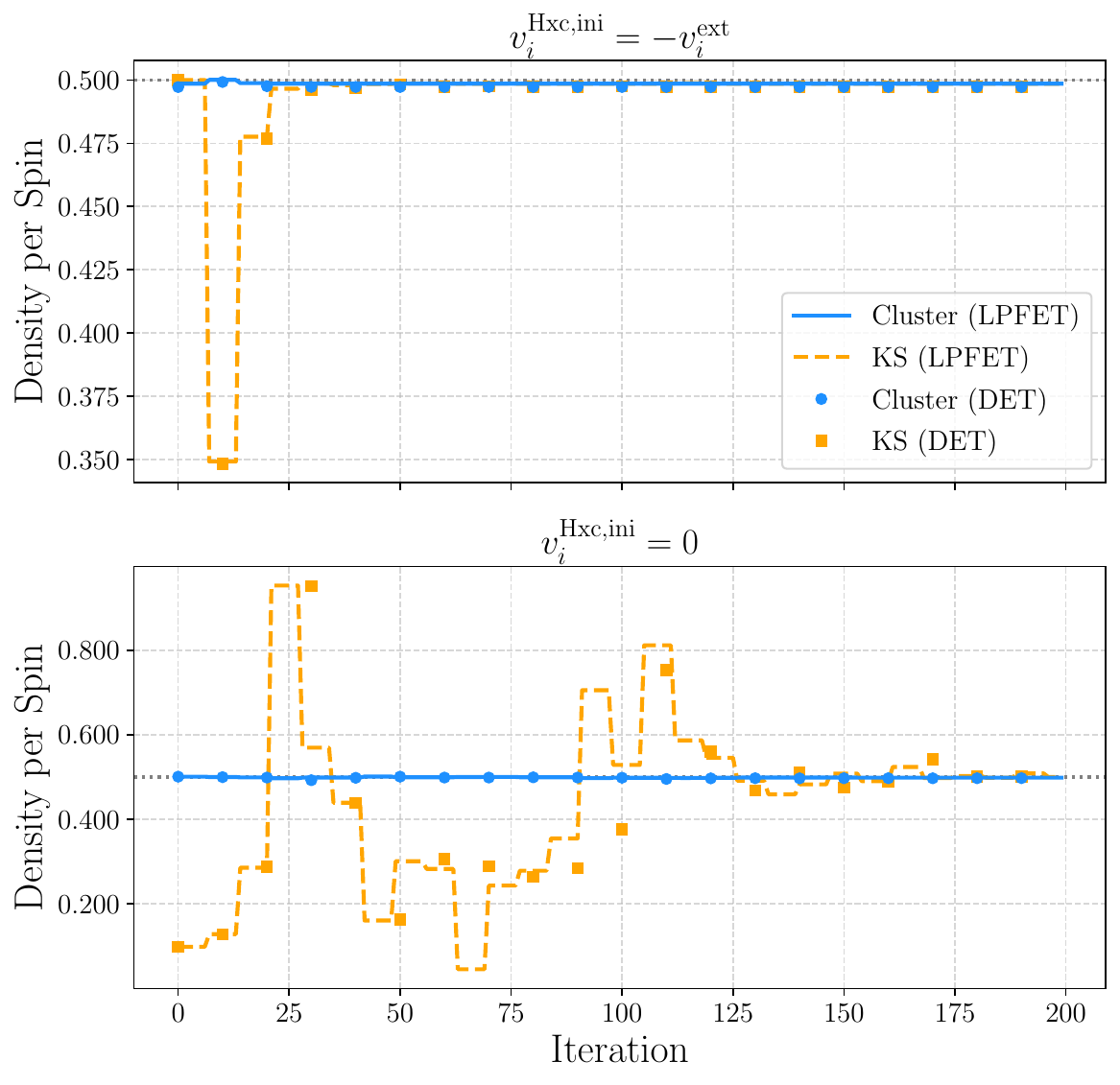}
    \caption{
    \rev{Convergence of LPFET densities on site 2 of the 6-site Hubbard ring in the strongly correlated $U/t=30$ regime, starting from different initial guesses: ${\bf v}^{\rm Hxc} = -{\bf v}^{\text{ext}}$ (top panel) and ${\bf v}^{\rm Hxc} = {\bf 0}$ (bottom panel). A tight threshold of $10^{-9}$ is used so that the KS and embedding cluster densities match at convergence. See text for further details.}
    }
    \label{Fig:Density_Convergence}
\end{figure}
For completeness, we tested the robustness of LPFET and DET algorithms in this regime, by imposing a very tight threshold ($10^{-9}$) on the density convergence. Results are shown for site 2 in Fig.~\ref{Fig:Density_Convergence}.
\rev{As readily seen, LPFET has the same convergence profile as DET in this strongly correlated regime.
Interestingly, while the cluster density
remains almost constant during the entire process,
the occupation of the KS system can fluctuate heavily,
showing that the impact of the impurity chemical potential
on the cluster density is much more subtle than the
impact of the Hxc potential on the KS system.
The latter will therefore dictate the convergence of the
LPFET and DET methods.
Note that the convergence is highly sensitive to the
initial values of the Hxc potential,
both in LPFET and DET. Indeed,
according to the top and bottom panels of Fig.~\ref{Fig:Density_Convergence},
taking as initial values $v^{\rm Hxc,ini}_i = - v^{\rm ext}_i$
leads to a KS density that is already very close to the cluster one, thus drastically reducing the fluctuations of the KS density during convergence compared
to taking $v^{\rm Hxc,ini}_i = 0, ~\forall i$.
In addition, while in DET the Hxc potential differences (with respect to some reference site) and the impurity chemical potential are treated as independent variables [see Eq.~(\ref{eq:DMET_emb_cluster_wf_opti})],
LPFET uses all the Hxc potential values (not up to a constant anymore) as basic variables and explicitly relate them to the impurity chemical potential within each embedding cluster (see Eqs.~(\ref{eq:LPFET_cluster_wf_variational_calc}) and (\ref{eq:ind_LPFET_chem_pot})).
Although such a relation is highly nonlinear, it is clear from Fig.~\ref{Fig:Density_Convergence} that, like DET, the LPFET algorithm still manages to converge.}\\
\begin{figure}
    \centering
    \includegraphics[width=1\linewidth]{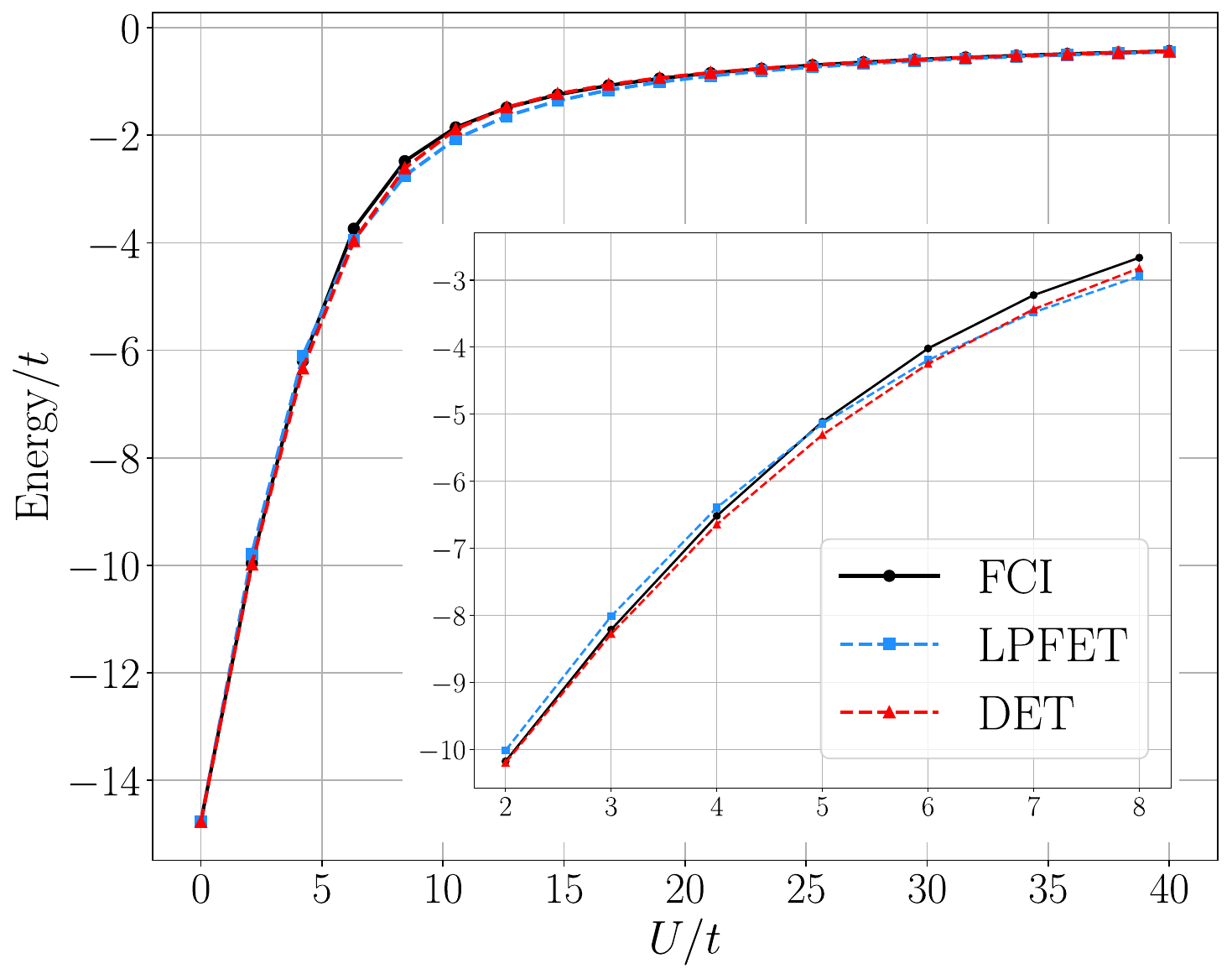}
    \caption{Total energy of a (non-uniform and half-filled) six-site Hubbard ring evaluated as a function of the interaction strength $U/t$ from LPFET and DET calculations. Comparison is made with FCI. See text for further details.}
    \label{Energy_vs_U}
\end{figure}

\begin{figure}
    \centering
    \includegraphics[width=1.4\linewidth]{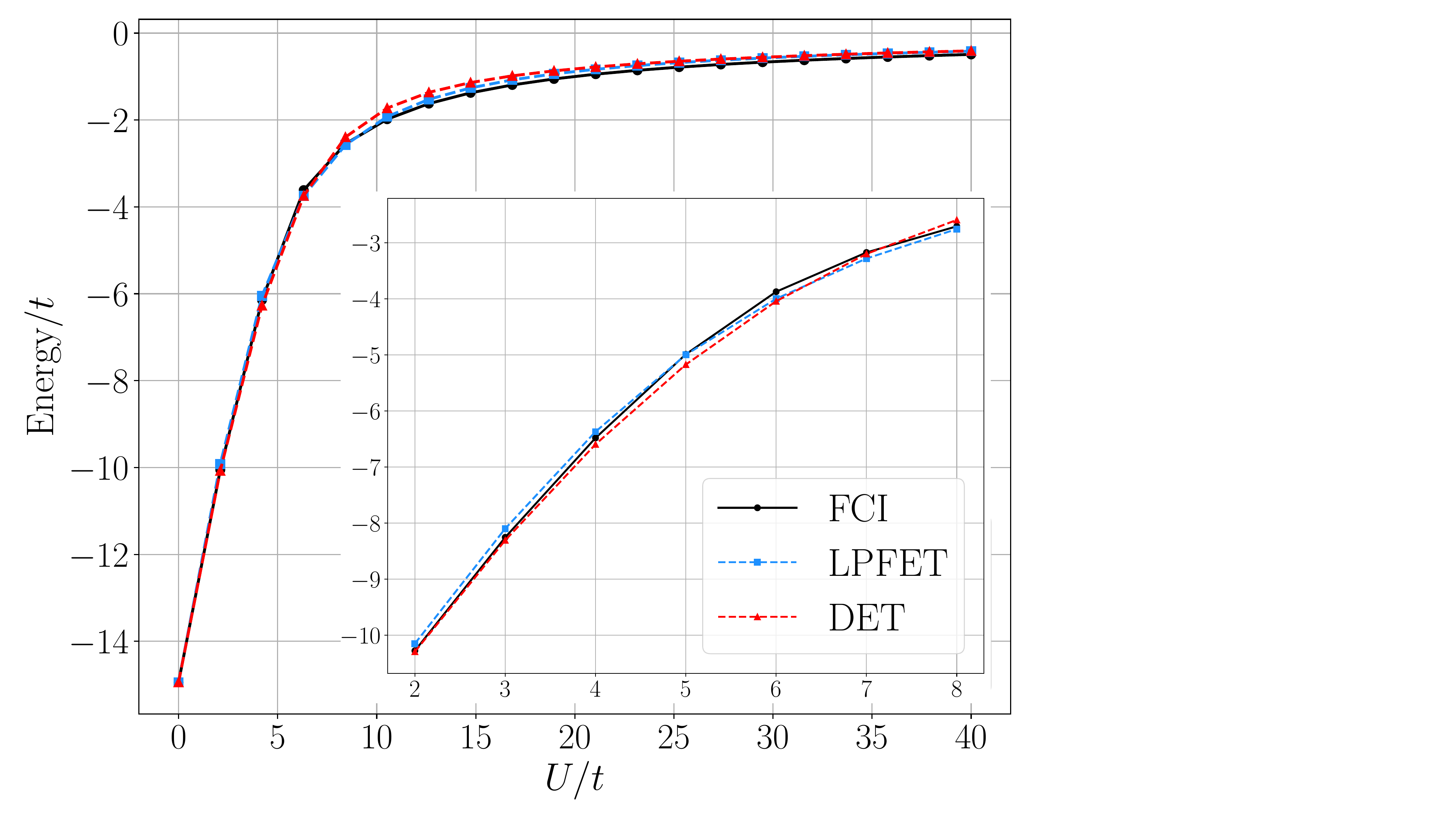}
    \caption{Total energy of a (non-uniform and half-filled) six-site Hubbard ladder evaluated as a function of the interaction strength $U/t$ from LPFET and DET calculations. Comparison is made with FCI. }
    \label{Energy_vs_U_2D}
\end{figure}

Let us now turn our attention to evaluating the total energy of the system. Accurate energy prediction is a critical validation step~\cite{nusspickel2023effective}. At convergence, correlated one- and two-electron RDMs can be evaluated within each embedding cluster. Combining these RDMs to reconstruct total (ideally $N$-representable) full-size RDMs is a key and still challenging task~\cite{nusspickel2023effective} in quantum embedding theory. It will not be addressed in the present work. Instead, we use the (standard) energy expression of Wouters {\it et al.} (see Eq.~(28) in Ref.~\citenum{wouters2016practical}), where a so-called ``democratic partitioning'' is used. If we focus on evaluating the expectation value for the per-site kinetic energy operator (see Eq.~(\ref{eq:per-site_kinetic_ener_op_for_emb})), for example, we compute the one-electron RDM as follows,  
\be
\left\langle\hat{c}_{i\sigma}^\dagger\hat{c}_{j\sigma}+\hat{c}_{j\sigma}^\dagger\hat{c}_{i\sigma}\right\rangle\underset{j>i}{\approx} 
\left\langle\hat{c}_{i\sigma}^\dagger\hat{c}_{j\sigma}\right\rangle_{\Psi^{(i)}}+\left\langle\hat{c}_{j\sigma}^\dagger\hat{c}_{i\sigma}\right\rangle_{\Psi^{(j)}}.
\ee
The total energy is plotted as a function of $U/t$ in Fig.~\ref{Energy_vs_U}. 
\rev{The overall performance of LPFET is relatively good. Nevertheless, in contrast to DET, LPFET gives too high energies in the weakly $2\leq U/t\leq 4$ correlated regime (see the inset in Fig.~\ref{Energy_vs_U}) while, in the stronger $8\leq U/t\leq 15$ correlation regime, it overestimates the correlation energy. In the vicinity of $U/t=5$, LPFET is highly accurate and outperforms DET, which is probably due to error cancellations. It is important to recall that the above-mentioned democratic evaluation of RDMs is actually an arbitrary (but convenient) choice. Questioning this choice is even more relevant in the context of LPFET, for which impurity chemical potentials are evaluated for each embedded fragment orbital, specifically, unlike in DET. There is no obvious solution to this problem~\cite{nusspickel2023effective} that we plan to investigate further in the future. 
In this task, we should ideally reach a variational construction of the total energy. Regarding this point, when DET performs better than LPFET, it still gives energies that are systematically lower than the exact FCI energy. Interestingly, in the weakly $2\leq U/t\leq 4$ correlated regime, where LPFET gives inaccurate density profiles, the LPFET energy is an upper bound to the exact FCI one.\\ 
}

\rev{
For completeness, we show in Fig.~\ref{Energy_vs_U_2D} the energy plot for the 2D Hubbard ladder model introduced earlier in this section. Except the fact that the correlation energy is not overestimated anymore in the strong $8\leq U/t\leq 15$ correlation regime, at both LPFET and DET levels, no major difference with the (1D) Hubbard ring can be seen in other correlation regimes (see Fig.~\ref{Energy_vs_U}, for comparison).  
}

\subsection{Ab initio hydrogen chain}\label{Hydrogen_chain}

\begin{figure}
    \centering
    \includegraphics[width=1\linewidth]{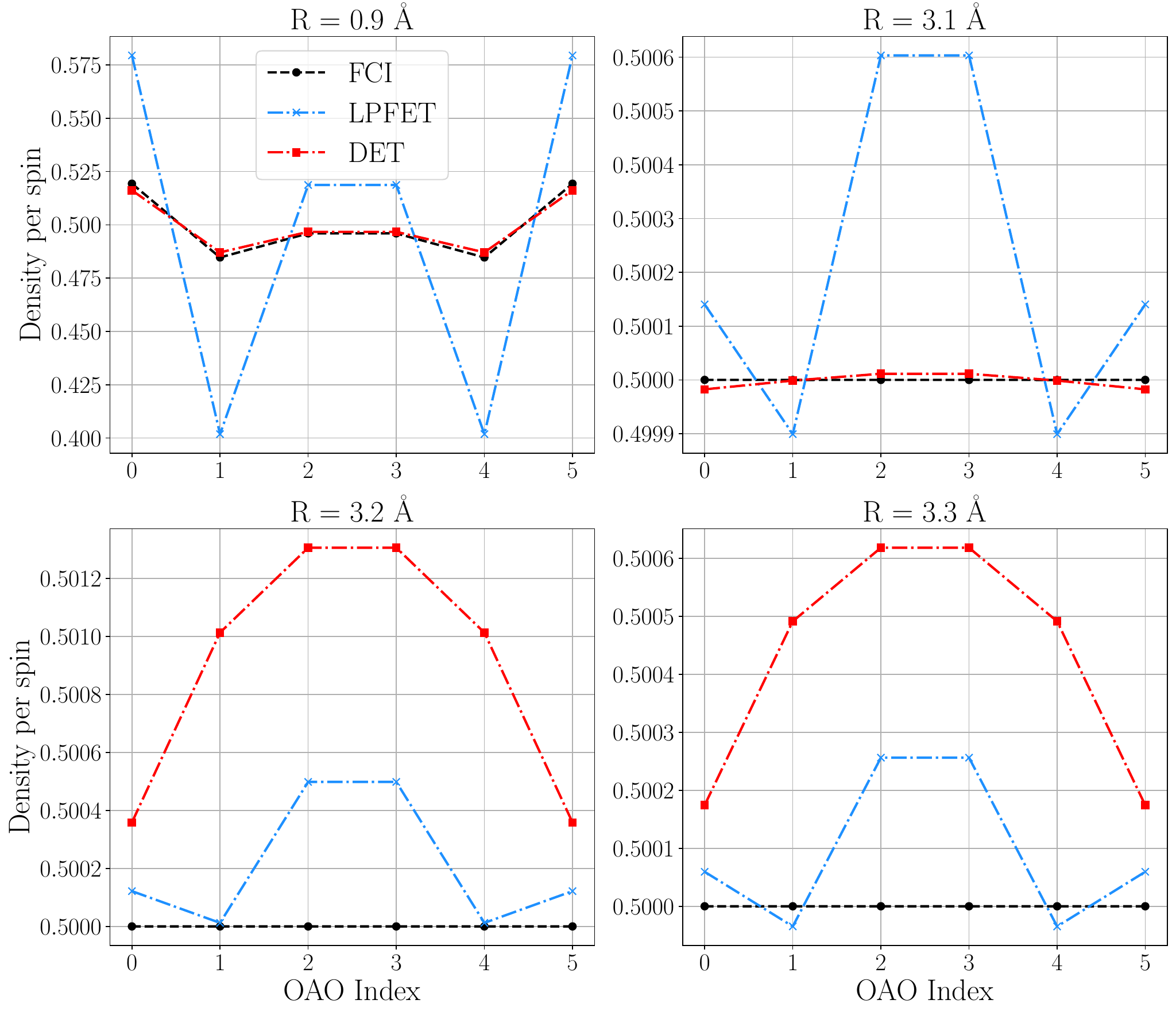}
    \caption{Self-consistently converged LPFET and DET density profiles (\ie, occupations of OAOs in this context) obtained for a chain of six equispaced hydrogen atoms and various bond distances. Comparison is made with FCI. See text for further details.}
    \label{Density_Hchain}
\end{figure}
\begin{figure}
    \centering
    \includegraphics[width=1\linewidth]{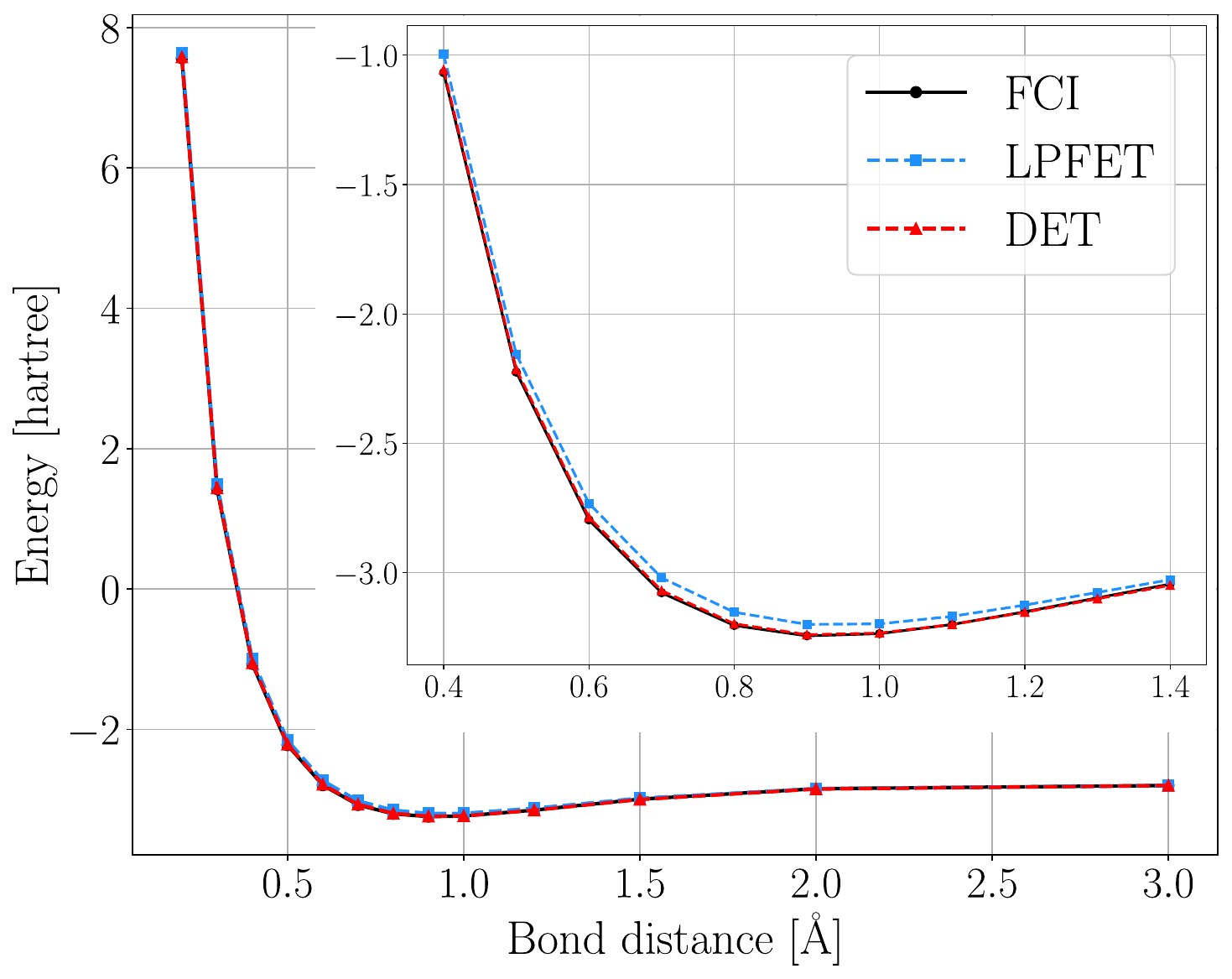}
    \caption{Potential energy curves computed from LPFET and DET calculations for a chain of six equispaced hydrogen atoms. Comparison is made with FCI. As shown in the inset, the equilibrium distance is around $R=0.9$ {\AA} (the energy equals -3.27 Hartree in this case). See text for further details.}
    \label{Energy_focuse_Hchain}
\end{figure}

LPFET is, like DMET or DET, applicable not only to lattice models but also to {\it ab initio} quantum chemical Hamiltonians. In this section, we discuss the results obtained for a prototypical chain of six equispaced hydrogen atoms (see Sec.~\ref{sec:comput_details} for the computational details). 
Varying the bond distance allows us to explore various correlation regimes. We start with the analysis of the self-consistently converged density profiles (\ie, the occupations of L\"{o}wdin's OAOs in this context) that are shown in Fig.~\ref{Density_Hchain}. 
\rev{As readily seen on
the top-left panel,
which corresponds to the equilibrium distance
of the hydrogen chain and to the weakly correlated regime,
LPFET does not provide very accurate orbital occupations
in contrast to DET, in analogy to the weakly correlated regime of the Hubbard model.
This trend is true up to the dissociation of the
hydrogen chain ($R < 3.1$~\AA),
and tends to reverse for $R > 3.1$~{\AA} (see $R=3.2$~{\AA} and
$R = 3.3$~{\AA} in the bottom panels of Fig.~\ref{Density_Hchain}),
thus showing that LPFET is indeed accurate in the strongly correlated regime.
Note that these
subtle differences mainly come from the convergence of both methods,
where the density constraint in Eq.~(\ref{eq:LPFET_dens_const_sc_loop})
is not perfectly achieved (\ie, with an error
of the order of $10^{-3}$--$10^{-4}$ instead of $10^{-9}$ in the weakly correlated regime).
This convergence issue
is known in D(M)ET
and originates from
the low-level system calculation,
for which a single Slater determinant is not sufficient anymore to describe the system at dissociation.
Indeed, for $R>3.3$~\AA,
the highest occupied orbitals get degenerate
such that a description beyond closed-shell single Slater determinant is required,
such as proposed in Ref.~\citenum{cernatic2024fragment}
by using a Gross--Oliveira--Kohn ensemble
or in Ref.~\citenum{cui2025ab} by using a finite-temperature smearing.
LPFET is not exempt from the problem,
and relies on highly non-linear self-consistent equations that could make the convergence even
harder than in DET, at least in principle.
Nevertheless, within the systems studied in this work,
we observed similar convergence 
between LPFET and DET (see the discussion of Fig.~\ref{Fig:Density_Convergence}).\\
}
\rev{
The potential energy curves obtained with LPFET and DET are shown in Fig.~\ref{Energy_focuse_Hchain}. Like for the Hubbard models discussed previously, the standard ``democratic'' evaluation of RDMs, as introduced in the context of DMET~\cite{wouters2016practical}, has been used. Both methods perform well except LPFET which, close to equilibrium, underestimates the correlation energy. This feature, that we also observed in the weakly correlated Hubbard ring and ladder models, can be related to the incorrect density profiles that LPFET gives at convergence.\\   
}
\rev{As a final remark, unlike in the case of Hubbard models (where the poor performance of LPFET in the weakly correlated regime originates from the fact that the method derives from the locally and strongly correlated regime), LPFET suffers from another issue which is critical for {\it ab initio} systems, namely the inadequate description of longer-range interactions. In this respect, the combination  of a non-local (in the localized orbitals representation) exact exchange potential with a local correlation potential, which would be the local potential to be optimized in such a generalized LPFET, is highly desirable. From this (hopefully) better starting point, we may also want to recover longer-range correlations. Combining the LPFET formalism with the random phase approximation, as already explored in the context of DMET~\cite{scott2021extending}, is an appealing strategy. This is left for future work.
}

\section{Conclusion} \label{Con}

The quantum embedding of localized orbitals, as implemented in standard DMET calculations, has been revisited from the perspective of (lattice) DFT. An in-principle exact and general (\ie, valid for both model lattice and {\it ab initio} quantum chemical Hamiltonians) formulation of the theory, where the occupations of the (arbitrarily constructed) localized orbitals play the role of the density, has been derived. Thus, we extend and generalize a theory (LPFET) that was originally intended to describe the thermodynamic limit of the uniform one-dimensional Hubbard model~\cite{sekaran2022local}. By doing so, we also provide a more rigorous framework for approaches that combine DFT and DMET formalisms, such as DET~\cite{bulik2014density} or SDE~\cite{mordovina2019self}. Let us stress that, in this context, ``exact'' means FCI, as the theory is written in an orbital basis of arbitrary but finite size. While presented in detail for single-orbital fragments, the derivation can be generalized to multiple-orbital embeddings. In this context, the full-size (so-called ``low-level'' in DMET) calculation is becoming exact, in the sense of (lattice) KS-DFT, as it is expected to reproduce the exact density profile. By exploiting the present density-functional formalism, we could establish an exact relation between the Hxc potential of the KS system and the ({\it a priori} fragment-dependent) embedding potential that is introduced within each embedding cluster (see Eq.~(\ref{eq:LPFET_cluster_wf_variational_calc})), individually, and is referred to as impurity chemical potential. The latter potential ensures that the embedding cluster reproduces exactly, like the KS system, the local density. By applying well-identified density-functional approximations, which can be justified when electron correlation is strongly local, the practical LPFET approach naturally emerges from the theory. While DET uses, in the common IB setting, a global (\ie, fragment-independent) chemical potential as embedding potential, the impurity chemical potential of LPFET is expressed explicitly in terms of the (local, hence the name of the theory) Hxc potential and the bath orbital of the embedding cluster (as readily seen from the key Eq.~(\ref{eq:simplified_imp_chem_pot_LPFET})), thus making it fragment-dependent. Note that, in its practical implementation, LPFET performs no KS inversion within the embedding clusters, unlike SDE~\cite{mordovina2019self}. Instead, all Hxc potential values are updated at once, by comparing the KS density profile with that deduced from all local densities. The latter are computed individually at the correlated level within each embedding cluster. Convergence is reached when the two density profiles match.\\
Illustrative calculations have been performed on prototypical six-electron systems: non-uniform Hubbard ring and ladder models, and a chain of equispaced hydrogen atoms. They confirm the relevance of using a fragment-dependent embedding potential in place of a global one, when electron correlation becomes strong, a regime where LPFET finds its theoretical justification. Indeed, in this case, LPFET outperforms DET, providing systematically highly accurate density profiles. In the weakly correlated regime, however, LPFET is less accurate than DET. A possible rigorous remedy could be the generalized KS formalism~\cite{seidl1996generalized}, where a non-local (in the lattice representation) KS-orbital-dependent Hx potential would be combined with a (to-be-determined, self-consistently) local correlation potential. In the light of the DET results, we can expect a better performance of LPFET if the latter potential only contributes to the impurity chemical potential. This option will be investigated further in future work.\\
\rev{Total energies have been computed following the standard ``democratic partitioning'' in the one- and two-electron RDMs evaluation. Even though the procedure is not variational, we note that, when LPFET performs poorly in terms of density profile (in the weakly correlated regime), it provides an upper bound to the FCI energy.} The democratic partitioning is questionable in the case of LPFET, as the method treats each embedding cluster individually, through its non-global evaluation of the impurity chemical potentials. 
Even though this is a difficult problem to solve~\cite{nusspickel2023effective}, as it relates more generally to the $N$-representability of the computed RDMs and the variationality of the total energy in embedding calculations, practical solutions that combine the (non-orthogonal) correlated wavefunctions of each cluster should be further explored, with possible useful applications in quantum computing for which quantum embedding techniques appear more and more promising~\cite{rubin2016hybrid,ma2020quantum,li2022toward,otten2022localized,greene2022modelling,cao2023ab,d2024fragment,koridon2025fragpt2,erakovic2025high,verma2025multireference,weisburn2025multiscale}.
Work is currently in progress in this direction. Finally, even though the theory has been derived for pure ground states, the extension to density-functional ensembles of ground and excited states~\cite{Cernatic_2021,gould2025ensemblization} should be considered as a natural continuation, thus placing earlier work along those lines~\cite{cernatic2024fragment} in a more rigorous and broader perspective.     



\section{Acknowledgments}

This work of the Interdisciplinary Thematic Institute QMat, as part of the ITI 2021‐2028 program of the University of Strasbourg, CNRS and Inserm, was supported by IdEx Unistra (ANR 10 IDEX 0002), and by SFRI STRAT’US project (ANR 20 SFRI 0012) and EUR QMAT ANR‐17‐EURE‐0024 under the framework of the French Investments for the Future Program. This work benefited also from State support managed by the National Research Agency under the France 2030 program, referenced by ANR‐22‐CMAS‐0001 as well as ANR-23-PETQ-0006. The authors are grateful to Dr.~Saad Yalouz for stimulating discussions and for making the QuantNBody package \cite{Yalouz2022QuantNBody} available to us. The authors also thank Dr. M. Sauban\`{e}re for his useful comments on this work.

\appendix

\section{Exact and approximate density-functional impurity chemical potentials for a general partially-interacting bath}\label{appendix:relations_chem_pots_partial-int_bath}

In the spirit of site occupation embedding theory (SOET)~\cite{Fromager_2015, senjean2017local,
    senjean2018site,
    senjean2018multiple,
    senjean2019projected}, let us consider the following full-size partially-interacting Hamiltonian whose ground state reproduces the density profile $\undern$:
\be\label{eq:partially_int_dens_fun_Hamil}
\begin{split}
\hat{H}^{(i)\mathcal{N}}(\undern)
&=\hat{T}+\sum_{k}v_k^{\rm s}(\undern)\hat{n}_k
\\
&\quad
+\left(\hat{U}^{i+\mathcal{N}}_{i}-\sum_k\bar{v}_k^{(i)\mathcal{N}}(\undern)\hat{n}_k\right).
\end{split}
\ee
In such a fictitious Hamiltonian, electronic interactions are ``turned on'' from the impurity $\chi_i$ up to the site $\chi_{i+\mathcal{N}}$ (in the sense of circular permutations), \ie,  
\be\label{eqapp:partial_electronic_rep}
\hat{U}^{i+\mathcal{N}}_{i}\equiv \dfrac{1}{2}\sum_{i\leq k,l,m,n, \leq i+\mathcal{N}}\sum_{\sigma,\sigma'}\langle kl\vert mn\rangle\hat{c}_{k\sigma}^\dagger\hat{c}_{l\sigma'}^\dagger\hat{c}_{n\sigma'}\hat{c}_{m\sigma}, \nonumber \\
\ee
where the integer $\mathcal{N}$ indicates the range of the partial interaction. By construction, minus the local potential $\left\{\bar{v}_k^{(i)\mathcal{N}}(\undern)\right\}$ systematically corrects the KS potential such that the ground-state density of $\hat{H}^{(i)\mathcal{N}}(\undern)$ does match $\undern$. While other (so-called) adiabatic connection paths could have been considered (the present one is obtained by varying $\mathcal{N}$), we are mostly interested in the two limiting cases, namely the non-interacting bath (NIB), for which 
\be\label{eqapp:range_int_NIB_case}
\mathcal{N}\overset{\rm NIB}{=}0
\ee
and
\be\label{eq:app_NIB_partial_interaction}
\hat{U}^{i+\mathcal{N}}_{i}\overset{\rm NIB}{\equiv}\hat{U}^{(i)},
\ee
and the fully-interacting bath (IB), where $\mathcal{N}+1$ equals the total number of sites in the full system, thus leading to
\be
\hat{U}^{i+\mathcal{N}}_{i}\overset{\rm IB}{\equiv}\hat{U}.
\ee
Note that, in this particular case, 
\be\label{eqapp:comp_dens_pot_fully_IB_case}
\bar{v}_k^{(i)\mathcal{N}}(\undern)\overset{\rm IB}{=}v_k^{\rm Hxc}(\undern),\;\forall i.
\ee
Returning to the general case, the resulting Schr{\"o}dinger equation for the embedding cluster reads 
\be\label{eqapp:part-IB_cluster_eq}
\hat{\mathcal{H}}^{(i)\mathcal{N}}(\undern)\myket{\Psi^{(i)\mathcal{N}}(\undern)}=\mathcal{E}^{(i)\mathcal{N}}(\undern)\myket{\Psi^{(i)\mathcal{N}}(\undern)},
\ee
where
\be\label{eqapp:dens_func_embb_cluster_Hamil_pIB}
\hat{\mathcal{H}}^{(i)\mathcal{N}}(\undern)=\hat{P}^{(i)}(\undern)\hat{H}^{(i)\mathcal{N}}(\undern)\hat{P}^{(i)}(\undern)-\mu^{(i)\mathcal{N}}_{\rm imp}(\undern)\hat{n}_i. \nonumber \\
\ee
Note the introduction, after projection, of the ($\mathcal{N}$-dependent in this case) impurity chemical potential $\mu^{(i)\mathcal{N}}_{\rm imp}(\undern)$. Its role is to ensure that the embedding cluster and its KS analog fulfill the same local density constraint, \ie,
\be\label{eqapp:part-int-bath_dens_constraint}
\innerop{\Psi^{(i)\mathcal{N}}(\undern)}{\hat{n}_i}{\Psi^{(i)\mathcal{N}}(\undern)}=n_i=\innerop{\Phi^{(i)}(\undern)}{\hat{n}_i}{\Phi^{(i)}(\undern)}. \nonumber \\
\ee
For convenience, we decompose the density-functional embedding cluster Hamiltonian as follows (see Eqs.~(\ref{eq:partially_int_dens_fun_Hamil}), (\ref{eqapp:dens_func_embb_cluster_Hamil_pIB}), and (\ref{eq:KS_cluster_Hamil_full_proj})),
\be\begin{split}\label{eqapp:partially_int_dens_func_cluster_Hamilt}
\hat{\mathcal{H}}^{(i)\mathcal{N}}(\undern)&\equiv\hat{\mathcal{H}}_{\rm s}^{(i)}(\undern)+\hat{U}^{(i)}-\mu^{(i)\mathcal{N}}_{\rm imp}(\undern)\hat{n}_i
\\
&\quad
+\hat{P}^{(i)}(\undern)\Delta \hat{U}_{i}^{i+\mathcal{N}}\hat{P}^{(i)}(\undern)
\\
&\quad
-\hat{P}^{(i)}(\undern)\left(\sum_k\bar{v}_k^{(i)\mathcal{N}}(\undern)\hat{n}_k\right)\hat{P}^{(i)}(\undern),
\end{split}
\ee
where we separate the electronic repulsion $\hat{U}^{(i)}$ on the impurity from the interactions beyond the impurity (up to range $\mathcal{N}$) that we denote $\Delta \hat{U}_{i}^{i+\mathcal{N}}=\hat{U}^{i+\mathcal{N}}_{i}-\hat{U}^{(i)}$. Through the projection operator $\hat{P}^{(i)}(\undern)$ the latter interactions are described exactly between electrons within the cluster and at the mean-field level of approximation between core and cluster electrons. Note that in the two limiting cases of interest (NIB and IB) we have  
\be
\Delta \hat{U}_{i}^{i+\mathcal{N}} \overset{\rm NIB}{\equiv} 0,
\ee
and
\be\label{eqappendix:IB_int_beyond_imp}
\Delta \hat{U}_{i}^{i+\mathcal{N}} \overset{\rm IB}{\equiv} \hat{U} - \hat{U}^{(i)},
\ee
respectively. If we remove the core contributions to the last term on the right-hand side of Eq.~(\ref{eqapp:partially_int_dens_func_cluster_Hamilt}) and move, by a constant shift, the bath contributions (there are no impurity-bath coupling terms) to the impurity, we obtain the following alternative expression:
\be\begin{split}\label{eqapp:partially_int_dens_func_cluster_Hamilt_simp}
\hat{\mathcal{H}}^{(i)\mathcal{N}}(\undern)
&
\equiv\hat{\mathcal{H}}_{\rm s}^{(i)}(\undern)+\hat{U}^{(i)}-\mu^{(i)\mathcal{N}}_{\rm imp}(\undern)\hat{n}_i
\\
&
+\hat{P}^{(i)}(\undern)\Delta \hat{U}_{i}^{i+\mathcal{N}}\hat{P}^{(i)}(\undern)
\\
&-\left(\bar{v}_i^{(i)\mathcal{N}}(\undern)-\sum_k\inner{b^{(i)}(\undern)}{\chi_k}^2\bar{v}_k^{(i)\mathcal{N}}(\undern)\right)\hat{n}_i.
\end{split}
\ee
As a generalization (for a partially-IB) of Eq.~(\ref{eq:clusterized_HK_func_exp}), we consider the following decomposition of the exact functional 
\be\label{eqapp:clusterized_HK_func_exp_part-int-bath}
\begin{split}
F(\undern)&=\sum_i\innerop{\Psi^{(i)\mathcal{N}}(\undern)}{\hat{t}^{(i)} + \hat{U}^{(i)}}{\Psi^{(i)\mathcal{N}}(\undern)}
+\bar{E}^{\mathcal{N}}_{\rm Hxc}(\undern),
\end{split}
 \nonumber \\
\ee
thus leading, when comparison is made with the KS decomposition of Eq.~(\ref{eq:KS_decomp_from_clusters}), to 
\be
\begin{split}
E_{\rm Hxc}(\undern)-\bar{E}^{\mathcal{N}}_{\rm Hxc}(\undern)&=\sum_i\innerop{\Psi^{(i)\mathcal{N}}(\undern)}{\hat{t}^{(i)} + \hat{U}^{(i)}}{\Psi^{(i)\mathcal{N}}(\undern)}
\\
&\quad
-\sum_i\innerop{\Phi^{(i)}(\undern)}{\hat{t}^{(i)}}{\Phi^{(i)}(\undern)}.
\end{split}
\ee
As shown in the rest of the appendix, the above equation will provide, once it has been differentiated with respect to the density, an exact relation between the impurity chemical potentials $\mu^{(i)\mathcal{N}}_{\rm imp}(\undern)$ and the Hxc potential. Indeed, by equalizing the density-functional derivatives as follows,  
\be\label{eqapp:deriv_diff_E_Hxc_E_Hxc_comp}
\begin{split}
&\dfrac{\partial}{\partial n_j}
\left(E_{\rm Hxc}(\undern)-\bar{E}^{\mathcal{N}}_{\rm Hxc}(\undern)\right)\\
&
\underset{j\neq r}{=}2\sum_i\innerop{\dfrac{\partial \Psi^{(i)\mathcal{N}}(\undern)}{\partial n_j}}{\hat{t}^{(i)} + \hat{U}^{(i)}}{\Psi^{(i)\mathcal{N}}(\undern)}
\\
&\quad
-2\sum_i\innerop{\dfrac{\partial\Phi^{(i)}(\undern)}{\partial n_j}}{\hat{t}^{(i)}}{\Phi^{(i)}(\undern)},
\end{split}
\ee
and connecting both interacting and KS embedding cluster contributions (first and second terms on the right-hand side of Eq.~(\ref{eqapp:deriv_diff_E_Hxc_E_Hxc_comp}), respectively) to the associated Hamiltonians $\hat{\mathcal{H}}^{(i)\mathcal{N}}(\undern)$ and $\hat{\mathcal{H}}_{\rm s}^{(i)}(\undern)$, we obtain the more explicit and informative relation (see Eqs.~(\ref{eqapp:partially_int_dens_func_cluster_Hamilt_simp}), (\ref{eq:Hxc_pot_func_deriv}), (\ref{eq:non_int_KS_cluster_dens_func_Hamil_all_on_imp}),(\ref{eq:dens_residual_site}), and (\ref{eq:tau_op_def})),
\be\label{eqapp:developt_dens_func_deriv_nj}
\begin{split}
&
v^{\rm Hxc}_j(\undern)-v^{\rm Hxc}_r(\undern)-\dfrac{\partial \bar{E}^{\mathcal{N}}_{\rm Hxc}(\undern)}{\partial n_j}
\\
&\underset{j\neq r}{=}2\sum_i\innerop{\dfrac{\partial \Psi^{(i)\mathcal{N}}(\undern)}{\partial n_j}}{\hat{\mathcal{H}}^{(i)\mathcal{N}}(\undern)}{\Psi^{(i)\mathcal{N}}(\undern)}
\\
&-2\sum_i\innerop{\dfrac{\partial\Phi^{(i)}(\undern)}{\partial n_j}}{\hat{\mathcal{H}}_{\rm s}^{(i)}(\undern)}{\Phi^{(i)}(\undern)}
\\
&
-2\sum_i\innerop{\dfrac{\partial \Psi^{(i)\mathcal{N}}(\undern)}{\partial n_j}}{\hat{\tau}^{(i)}(\undern)}{\Psi^{(i)\mathcal{N}}(\undern)}
\\
&+2\sum_i\innerop{\dfrac{\partial\Phi^{(i)}(\undern)}{\partial n_j}}{\hat{\tau}^{(i)}(\undern)}{\Phi^{(i)}(\undern)}
\\
&
-2\sum_i\innerop{\dfrac{\partial \Psi^{(i)\mathcal{N}}(\undern)}{\partial n_j}}{\hat{P}^{(i)}(\undern)\Delta \hat{U}_{i}^{i+\mathcal{N}}\hat{P}^{(i)}(\undern)}{\Psi^{(i)\mathcal{N}}(\undern)}
\\
&
+2\sum_i\left(\bar{v}_i^{(i)\mathcal{N}}(\undern)-\sum_k\inner{b^{(i)}(\undern)}{\chi_k}^2\bar{v}_k^{(i)\mathcal{N}}(\undern)+\mu^{(i)\mathcal{N}}_{\rm imp}(\undern)\right)
\\
&\times\innerop{\dfrac{\partial \Psi^{(i)\mathcal{N}}(\undern)}{\partial n_j}}{\hat{n}_i}{\Psi^{(i)\mathcal{N}}(\undern)}
\\
&-2\sum_i\left(v^{\rm s}_i(\undern)-\sum_{k\neq i}\inner{b^{(i)}(\undern)}{\chi_k}^2v^{\rm s}_k(\undern)\right)
\\
&\times \Bigg(\innerop{\dfrac{\partial \Psi^{(i)\mathcal{N}}(\undern)}{\partial n_j}}{\hat{n}_i}{\Psi^{(i)\mathcal{N}}(\undern)}
-\innerop{\dfrac{\partial \Phi^{(i)}(\undern)}{\partial n_j}}{\hat{n}_i}{\Phi^{(i)}(\undern)}\Bigg).
\end{split}
\ee
The two first terms on the right-hand side of Eq.~(\ref{eqapp:developt_dens_func_deriv_nj}) vanish because of the (partially) IB and KS embedding cluster equations (see Eqs.~(\ref{eqapp:part-IB_cluster_eq}) and (\ref{eq:KS_dens_func_cluster_wf_def}), respectively) while the last term also vanishes, because of the density constraint of Eq.~(\ref{eqapp:part-int-bath_dens_constraint}). If we now introduce the following density bi-functionals, 
\be
\tau^{\mathcal{N}}_{\rm c}(\undernu,\undern)=\sum_i\tau_{\rm c}^{(i)\mathcal{N}}(\undernu,\undern),
\ee
where (see Eq.~(\ref{eq:tau_op_def}))
\be
\begin{split}
\tau_{\rm c}^{(i)\mathcal{N}}(\undernu,\undern)&=\innerop{\Psi^{(i)\mathcal{N}}(\undern)}{\hat{\tau}^{(i)}(\undernu)}{\Psi^{(i)\mathcal{N}}(\undern)}
\\
&\quad
-\innerop{\Phi^{(i)}(\undern)}{\hat{\tau}^{(i)}(\undernu)}{\Phi^{(i)}(\undern)}
,
\end{split}
\ee
and
\be
\mathcal{U}^{\mathcal{N}}_{\rm Hxc}(\undernu,\undern)=\sum_i\mathcal{U}^{(i)\mathcal{N}}_{\rm Hxc}(\undernu,\undern)
,
\ee
where
\be
\begin{split}
&\mathcal{U}^{(i)\mathcal{N}}_{\rm Hxc}(\undernu,\undern)
=\innerop{\Psi^{(i)\mathcal{N}}(\undern)}{\hat{P}^{(i)}(\undernu)\Delta \hat{U}_{i}^{i+\mathcal{N}}\hat{P}^{(i)}(\undernu)}{\Psi^{(i)\mathcal{N}}(\undern)},
\end{split}
\nonumber \\
\ee
using the simplified density derivative expressions,
\be
2\innerop{\dfrac{\partial \Psi^{(i)\mathcal{N}}(\undern)}{\partial
n_j}}{\hat{n}_i}{\Psi^{(i)\mathcal{N}}(\undern)}=\dfrac{\partial
n_i}{\partial n_j}\overset{i\neq r}{\underset{j\neq r}{=}}\delta_{ij}
\ee
and (see Eq.~(\ref{eq:dens_residual_site}))
\be
2\innerop{\dfrac{\partial \Psi^{(r)\mathcal{N}}(\undern)}{\partial
n_j}}{\hat{n}_r}{\Psi^{(r)\mathcal{N}}(\undern)}=\dfrac{\partial
n_r}{\partial n_j}\underset{j\neq r}{=}-1,
\ee
leads to the following exact and general (partially-IB) relation: 
\be\label{eqapp:general_imp_chem_pots_relation_part-IB}
\begin{split}
-\mu^{(j)\mathcal{N}}_{\rm imp}(\undern)&\underset{j\neq r}{=}-\mu^{(r)\mathcal{N}}_{\rm imp}(\undern)-\left(v^{\rm Hxc}_j(\undern)-v^{\rm Hxc}_r(\undern)\right)
\\
&\quad
+\left(\bar{v}_j^{(j)\mathcal{N}}(\undern)-\bar{v}_r^{(r)\mathcal{N}}(\undern)\right)
\\
&\quad
-\sum_k\inner{b^{(j)}(\undern)}{\chi_k}^2\bar{v}_k^{(j)\mathcal{N}}(\undern)
\\
&\quad
+\sum_k\inner{b^{(r)}(\undern)}{\chi_k}^2\bar{v}_k^{(r)\mathcal{N}}(\undern)
\\
&\quad +\dfrac{\partial \bar{E}^{\mathcal{N}}_{\rm Hxc}(\undern)}{\partial n_j}
-\left.\dfrac{\partial \mathcal{U}^{\mathcal{N}}_{\rm Hxc}(\undernu,\undern)}{\partial n_j}\right|_{\undernu=\undern}
\\
&\quad
-\left.\dfrac{\partial \tau^{\mathcal{N}}_{\rm
c}(\undernu,\undern)}{\partial n_j}\right|_{\undernu=\undern}.
\end{split}
\ee
In the particular case of a fully IB (see Eqs.~(\ref{eqapp:comp_dens_pot_fully_IB_case}) and (\ref{eqappendix:IB_int_beyond_imp})), which is the most common situation
in quantum chemistry, the above relation can be further simplified and written as in
Eq.~(\ref{eq:exact_imp_chem_pots_relations}). In the following, we will take the density equal to the exact physical one ($\undern=\undern^{\Psi_0}$) and drop the dependence on $\undern$ (we denote $\mu^{(j)\mathcal{N}}_{\rm imp}:=\mu^{(j)\mathcal{N}}_{\rm imp}\left(\undern^{\Psi_0}\right)$, for example).\\

Let us now turn to practical approximations to the key Eq.~(\ref{eqapp:general_imp_chem_pots_relation_part-IB}). By analogy with the IB formulation of DMET~\cite{wouters2016practical}, we consider (for a given range $\mathcal{N}$) the following density-functional approximation,  
\be\label{eqapp:part-int-pot-corr_approx}
\begin{split}
\forall i,\;\;\bar{v}_j^{(i)\mathcal{N}}(\undern)&\underset{i\leq j\leq i+\mathcal{N}}{\approx} v_j^{\rm Hxc}(\undern)
\\
&\underset{j\notin[i,i+\mathcal{N}]}{\approx}0,
\end{split}
\ee
which is expected to be accurate when electron correlation is strongly local (see Eq.~(\ref{eq:partially_int_dens_fun_Hamil})). If, in addition, we neglect the following density-functional derivatives (see Eq.~(\ref{eq:phys_meaning_two-bi-funcs}) and the comment that follows regarding the physical meaning of those terms),
\be\label{eqapp:part-int_additional_DFAs}
\begin{split}
\left.\dfrac{\partial \tau_{\rm c}^{\mathcal{N}}(\undernu,\undern)}{\partial n_j}\right|_{\undernu=\undern}
&\approx
0\approx \left.\dfrac{\partial \mathcal{U}^{\mathcal{N}}_{\rm Hxc}(\undernu,\undern)}{\partial n_j}\right|_{\undernu=\undern},
\\
\dfrac{\partial \bar{E}^{\mathcal{N}}_{\rm Hxc}(\undern)}{\partial n_j}
&\approx 
0,
\end{split}
\ee
Eq.~(\ref{eqapp:general_imp_chem_pots_relation_part-IB}) becomes (for $\undern=\undern^{\Psi_0}$), 
\be\label{eqapp:approx_relation_imp_chem_pots-partIB}
\begin{split}
-\mu^{(j)\mathcal{N}}_{\rm imp}&\underset{j\neq r}{\approx}-\mu^{(r)\mathcal{N}}_{\rm imp}
\\
&\quad
-\sum_{j\leq k\leq j+\mathcal{N}}\inner{b^{(j)}}{\chi_k}^2{v}_k^{\rm Hxc}
\\
&\quad
+\sum_{r\leq k\leq r+\mathcal{N}}\inner{b^{(r)}}{\chi_k}^2{v}_k^{\rm Hxc}.
\end{split}
\ee
As the Hxc potential is unique up to a constant shift, we can adjust the constant such that, even in the exact theory and for a given range value $\mathcal{N}$,
\be\label{eqapp:unique_def_Hxc_pot_PI-bath}
\sum_{r\leq k\leq r+\mathcal{N}}\inner{b^{(r)}}{\chi_k}^2{v}_k^{\rm Hxc}:=\mu^{(r)\mathcal{N}}_{\rm imp},
\ee
where we recall that $\mu^{(r)\mathcal{N}}_{\rm imp}$ is uniquely defined from the exact local density constraint (see Eq.~(\ref{eqapp:part-int-bath_dens_constraint})). Eq.~(\ref{eqapp:approx_relation_imp_chem_pots-partIB}) immediately implies that the above equation actually holds (approximately) for {\it all} impurities, \ie, 
\be
\label{eqapp:approx_chem_pot_parIB}
\mu^{(i)\mathcal{N}}_{\rm imp}\approx\sum_{i\leq k\leq i+\mathcal{N}}\inner{b^{(i)}}{\chi_k}^2{v}_k^{\rm Hxc},\;\; \forall i. 
\ee
Let us finally inspect the partially-IB embedding cluster Hamiltonian, which reads as follows when the approximation of Eq.~(\ref{eqapp:part-int-pot-corr_approx}) is made [see Eqs.~(\ref{eqapp:dens_func_embb_cluster_Hamil_pIB}) and (\ref{eq:exact_KS_pot_on_site_i})],   
\be
\begin{split}
\hat{\mathcal{H}}^{(i)\mathcal{N}}&\approx \hat{P}^{(i)}\left(\hat{T}+\hat{U}_{i}^{i+\mathcal{N}}+\hat{V}^{\rm ext}\right)\hat{P}^{(i)}
\\
&\quad
+\hat{P}^{(i)}\left(\sum_{k\notin[i,i+\mathcal{N}]}v^{\rm Hxc}_k\hat{n}_k\right)\hat{P}^{(i)}-\mu^{(i)\mathcal{N}}_{\rm imp}\hat{n}_i.
\end{split}
\ee
By removing the scalar contributions from the core electrons and shifting the contribution from the bath to the impurity it comes, 
\be\label{eqapp:proj_Hxc_onto_non-int-sites_PIB}
\begin{split}
&\hat{P}^{(i)}\left(\sum_{k\notin[i,i+\mathcal{N}]}v^{\rm Hxc}_k\hat{n}_k\right)\hat{P}^{(i)}
\\
&
\equiv -\left(\sum_{k\notin[i,i+\mathcal{N}]}\inner{b^{(i)}}{\chi_k}^2{v}_k^{\rm Hxc}\right)\hat{n}_i
,
\end{split}
\ee
thus leading, according to Eq.~(\ref{eqapp:approx_chem_pot_parIB}), to the final and general approximate expression:
\be\label{eqapp:general-PI_bath_LPFET}
\begin{split}
\hat{\mathcal{H}}^{(i)\mathcal{N}}&\approx \hat{P}^{(i)}\left(\hat{T}+\hat{U}_{i}^{i+\mathcal{N}}+\hat{V}^{\rm ext}\right)\hat{P}^{(i)}
\\
&\quad
-\left(\sum_{k}\inner{b^{(i)}}{\chi_k}^2{v}_k^{\rm Hxc}\right)\hat{n}_i.  
\end{split}
\ee

\section{Clusterization kinetic correlation and beyond-the-impurity (but within-the-cluster) Hxc potential energies} \label{appendix:clusterization_kin_corr_energy}

Let us consider the ``off-diagonal'' projection of the kinetic energy operator onto the $i$th density-functional embedding cluster,
\be
\hat{\mathcal{T}}_{\rm off}^{(i)}(\undern):=\hat{P}^{(i)}(\undern)\hat{T}\hat{P}^{(i)}(\undern)
-\sum_\sigma\myexpval{\hat{T}}{b_\sigma^{(i)}(\undern)}\hat{b}^{(i)\dagger}_\sigma\hat{b}^{(i)}_\sigma,
\nonumber \\
\ee
which reads more explicitly as follows,
\be\label{eq:kin_ener_op_within_cluster_dens_func}
\hat{\mathcal{T}}_{\rm off}^{(i)}(\undern)=-\tilde{t}^{(i)}(\undern)\sum_\sigma\left(\hat{c}_{i\sigma}^\dagger\hat{b}^{(i)}_\sigma+{\rm H.c.}\right),
\ee
where the constant (scalar) contribution from the core electrons has been removed and $\tilde{t}^{(i)}(\undern)$ is an effective hopping term defined as
\be\label{eqapp:effective_hopping_cluster}
-\tilde{t}^{(i)}(\undern):=\sum_{j\neq i} h_{ij}\inner{\chi_j}{b^{(i)}(\undern)}.
\ee
We can now introduce the difference between the ``effective'' kinetic energy operator $\hat{\mathcal{T}}_{\rm off}^{(i)}(\undern)$ within the embedding cluster and its true physical (per-site) counterpart $\hat{t}^{(i)}$ (see Eq.~(\ref{eq:per-site_kinetic_ener_op_for_emb})),   
\be\label{eqapp:tau_off_op}
\hat{\tau}_{\rm off}^{(i)}(\undern):=
\hat{\mathcal{T}}_{\rm off}^{(i)}(\undern)-\hat{t}^{(i)},
\ee
that we refer to as {\it clusterization} kinetic energy operator. It is important to stress at this point that, according to the density constraints of Eq.~(\ref{eq:exact_dens_constraint_cluster}), and the fact that the number of electrons within each embedding cluster is fixed (and equal to 2, for a single-orbital fragment), the clusterization kinetic correlation energy introduced in Eq.~(\ref{eq:def_clusterization_kinetic_corr_ener_bifunc}) can be equivalently evaluated with $\hat{\tau}_{\rm off}^{(i)}(\undern)$ instead of the full operator $\hat{\tau}^{(i)}(\undern)$, which is defined in Eq.~(\ref{eq:tau_op_def}).\\

According to Eqs.~(\ref{eq:kin_ener_op_within_cluster_dens_func}) and (\ref{eq:per-site_kinetic_ener_op_for_emb}), the expression in Eq.~(\ref{eqapp:tau_off_op}) reads more explicitly as follows, 
\be
\begin{split}
\hat{\tau}_{\rm off}^{(i)}(\undern)&=-\tilde{t}^{(i)}(\undern)\sum_\sigma\left(\hat{c}_{i\sigma}^\dagger\hat{b}^{(i)}_\sigma+{\rm H.c.}\right)
\\
&\quad
- \sum_{j>i}h_{ij}\sum_{\sigma}\left(\hat{c}_{i\sigma}^\dagger\hat{c}_{j\sigma}+{\rm H.c.}\right),
\end{split}
\ee
which, since $\hat{b}^{(i)}_\sigma=\sum_{j\neq i}\inner{b^{(i)}(\undern)}{\chi_j}\hat{c}_{j\sigma}$,
leads (using real algebra) to the final expression:
\be
\label{eq:tau_operator_explicit_exp}
\begin{split}
\hat{\tau}_{\rm off}^{(i)}(\undern)
&=
-\tilde{t}^{(i)}(\undern)\sum_{j <i}\inner{b^{(i)}(\undern)}{\chi_j}\sum_{\sigma}\left(\hat{c}_{i\sigma}^\dagger\hat{c}_{j\sigma}+{\rm H.c.}\right)
\\
&-
\sum_{j>i}\left(\tilde{t}^{(i)}(\undern)\inner{b^{(i)}(\undern)}{\chi_j}+h_{ij}\right)\sum_{\sigma}\left(\hat{c}_{i\sigma}^\dagger\hat{c}_{j\sigma}+{\rm H.c.}\right).
\end{split}
\ee
From the first term (summation over $j<i$) in the right-hand side of Eq.~(\ref{eq:tau_operator_explicit_exp}) we immediately see that the clusterization kinetic energy operator is non-zero. It is the signature of the transformation that the per-site kinetic energy operator $\hat{t}^{(i)}$ undergoes (through a change of representation and projection) along the embedding process. Interestingly, if the fragment orbital $\chi_i$ were exclusively coupled by $\hat{T}$ to a single orbital $\chi_j$ (which is never the case in practice, otherwise the embedding would be pointless), with  $j>i$, meaning that the per-site kinetic energy operator $\hat{t}^{(i)}$ would simply be $ h_{ij}\sum_{\sigma}\left(\hat{c}_{i\sigma}^\dagger\hat{c}_{j\sigma}+{\rm H.c.}\right)$, according to Eq.~(\ref{eq:per-site_kinetic_ener_op_for_emb}), then $\chi_j\overset{j>i}{=}b^{(i)}(\undern)$ would play the role of the bath orbital, $\tilde{t}^{(i)}(\undern)$ would reduce to $-h_{ij}$ (see Eq.~(\ref{eqapp:effective_hopping_cluster})), and, consequently, $\hat{\tau}_{\rm off}^{(i)}(\undern)$ would simply vanish, as readily seen from Eq.~(\ref{eq:tau_operator_explicit_exp}).\\

\rev{Let us now focus on how the two density bi-functionals of Eqs.~(\ref{eq:def_clusterization_kinetic_corr_ener_bifunc}) and (\ref{eq:beyond_imp_within_cluster_Hxc_pot_ener_bi_func}) contribute to the total ``kinetic+electronic repulsion'' energy functional $F(\undern)$ [see Eq.~(\ref{eq:HK_func_def})].} By introducing, for a given impurity $\chi_i$, the following density-functional per-site kinetic correlation energy,
\be
t^{(i)}_{\rm c}(\undern)=\innerop{\Psi^{(i)}(\undern)}{\hat{t}^{(i)}}{\Psi^{(i)}(\undern)}-\innerop{\Phi^{(i)}(\undern)}{\hat{t}^{(i)}}{\Phi^{(i)}(\undern)}, \nonumber \\
\ee
the within-the-cluster kinetic correlation energy, 
\be
T^{(i)}_{\rm c}(\undern)=\innerop{\Psi^{(i)}(\undern)}{\hat{T}}{\Psi^{(i)}(\undern)}-\innerop{\Phi^{(i)}(\undern)}{\hat{T}}{\Phi^{(i)}(\undern)},\nonumber \\
\ee
and, finally, the within-the-cluster potential correlation energy, 
\be
\begin{split}
U^{(i)}_{\rm c}(\undern)
&=\innerop{\Psi^{(i)}(\undern)}{\hat{U}^{(i)}}{\Psi^{(i)}(\undern)}
\\
&\quad
-\innerop{\Phi^{(i)}(\undern)}{\hat{U}^{(i)}}{\Phi^{(i)}(\undern)},
\end{split}
\ee
and noticing that 
\be
\tau_{\rm c}^{(i)}(\undernu,\undern)\underset{\undernu=\undern}{=}T^{(i)}_{\rm c}(\undern)-t^{(i)}_{\rm c}(\undern),
\ee
$\Phi^{(i)}(\undern)=\hat{P}^{(i)}(\undern)\Phi^{(i)}(\undern)$, and $\Psi^{(i)}(\undern)=\hat{P}^{(i)}(\undern)\Psi^{(i)}(\undern)$, we can decompose the total kinetic and electronic interaction density-functional energy of the embedded wavefunction $\Psi^{(i)}(\undern)$ as follows,    
\be\begin{split}\label{eq:phys_meaning_two-bi-funcs}
&\innerop{\Phi^{(i)}(\undern)}{\hat{T}+\hat{U}^{(i)}}{\Phi^{(i)}(\undern)}
\\
&
+\left(t^{(i)}_{\rm c}(\undern)+U^{(i)}_{\rm c}(\undern)\right)+\tau_{\rm c}^{(i)}(\undernu,\undern)+
\mathcal{U}^{(i)}_{\rm Hxc}(\undernu,\undern)
\\
&
\underset{\undernu=\undern}{=}\innerop{\Psi^{(i)}(\undern)}{\hat{T}+\hat{U}}{\Psi^{(i)}(\undern)},
\end{split}
\ee
thus showing how the above-mentioned density bi-functionals contribute individually to the exact functional $F(\undern)$ [see the expression in Eq.~(\ref{eq:HK_func_def})], when a given embedding cluster (and the associated core) is used as reference. Note that, in order to approach the latter functional even further, excitations from the cluster and/or the core should be taken into account, ultimately. As an alternative to single-reference dynamical approaches~\cite{Booth24_PRL_Rigorous_Screened_Interactions}, multi-reference perturbation theory could be used for that purpose, for example. This is left for future work.

\section{Alternative but equivalent derivation of LPFET}\label{app:alternative_deriv_LPFET}

Our starting point will be the following {\it local} density-functional approximation to the Hxc potential,   
\be\label{eq:local_approx_within_cluster}
v_i^{\rm Hxc}=\left.\dfrac{\partial E_{\rm Hxc}({\bf n})}{\partial n_i}\right|_{{\bf n}=\undern^{\Psi_0}}\approx \left.\dfrac{\partial E^{(i)}_{\rm Hxc}(n_i)}{\partial n_i}\right|_{n_i=n_i^{\Psi_0}}, 
\ee
where the Hxc energy is evaluated (approximately) as follows, 
\be\label{eq:approx_local_evaluation_Hxc_ener}
\begin{split}
E^{(i)}_{\rm Hxc}(n_i)&=\innerop{\Psi^{(i)}(n_i)}{\hat{T}+\hat{U}}{\Psi^{(i)}(n_i)}
\\
&\quad
-\innerop{\Phi^{(i)}(n_i)}{\hat{T}}{\Phi^{(i)}(n_i)}
,
\end{split}
\ee
$\Psi^{(i)}(n_i)$ and $\Phi^{(i)}(n_i)$ being the normalized ground states of the interacting 
\be\label{eq:local_dens_func_int_cluster_Hamil}
\hat{\mathcal{H}}^{(i)}(n_i)=\hat{P}^{(i)}\hat{H}\hat{P}^{(i)}-\mu^{(i)}_{\rm imp}(n_i)\hat{n}_i
\ee
and non-interacting (we denote $\hat{h}^{\rm KS}=\hat{h}+\sum_k v_k^{\rm Hxc}\hat{n}_k$)
\be\label{eq:local_dens_func_KS_cluster_Hamil}
\hat{\mathcal{H}}_{\rm s}^{(i)}(n_i)=\hat{P}^{(i)}\hat{h}^{\rm KS}\hat{P}^{(i)}-\mu^{(i)}_{\rm s}(n_i)\hat{n}_i
\ee
{\it local} density-functional Hamiltonians, respectively. Let us stress at this point that, unlike in the exact theory of Sec.~\ref{sec:exact_DET}, the projection operator $\hat{P}^{(i)}\equiv \hat{P}^{(i)}\left(\undern^{\Psi_0}\right)$ onto the embedding cluster (or, equivalently, the bath orbital $b^{(i)}$) is assumed to be {\it fixed} in the approximation of Eq.~(\ref{eq:local_approx_within_cluster}). Only variations of the local density $n_i$ within the fixed embedding cluster are considered. The interacting $\mu^{(i)}_{\rm imp}(n_i)$ and non-interacting $\mu^{(i)}_{\rm s}(n_i)$ impurity chemical potentials are adjusted such that the following local density constraints are fulfilled:
\be\label{eq:local_int_dens_constraint}
\myexpval{\hat{n}_i}{\Psi^{(i)}(n_i)}\overset{!}{=}n_i
\ee
and
\be\label{eq:local_nonint_dens_constraint}
\myexpval{\hat{n}_i}{\Phi^{(i)}(n_i)}\overset{!}{=}n_i,
\ee
respectively. Note that, by construction (see Eqs.~(\ref{eq:exact_KS_pot_on_site_i}), (\ref{eq:KS_cluster_Hamil_full_proj}), (\ref{eq:KS_dens_func_cluster_wf_def}), and (\ref{eq:local_dens_mapping_KS})), 
\be\label{eq:no_imp_chem_pot_KS_system_exact_dens}
\mu^{(i)}_{\rm s}\left(n_i^{\Psi_0}\right)=0.
\ee
We should also stress that Eq.~(\ref{eq:local_approx_within_cluster}) provides an approximation to the exact Hxc potential that is, as readily seen from Eq.~(\ref{eq:approx_local_evaluation_Hxc_ener}), free from double counting of electron correlation. This observation echoes Sec.~\ref{sec:rationale_and_extensions_LPFET} regarding the IB formulation of DMET. We can now ``safely'' proceed with making Eq.~(\ref{eq:local_approx_within_cluster}) more explicit, which will lead ultimately to LPFET. For that purpose, we rewrite the Hxc density-functional energy of Eq.~(\ref{eq:approx_local_evaluation_Hxc_ener}) as follows, according to Eqs.~(\ref{eq:local_dens_func_int_cluster_Hamil}) and (\ref{eq:local_dens_func_KS_cluster_Hamil}),   
\be
\begin{split}
E^{(i)}_{\rm Hxc}(n_i)&=\innerop{\Psi^{(i)}(n_i)}{\hat{P}^{(i)}\left(\hat{T}+\hat{U}\right)\hat{P}^{(i)}}{\Psi^{(i)}(n_i)}
\\
&\quad
-\innerop{\Phi^{(i)}(n_i)}{\hat{P}^{(i)}\hat{T}\hat{P}^{(i)}}{\Phi^{(i)}(n_i)}
,
\end{split}
\ee
or, equivalently,
\be
\begin{split}\label{eq:cluster_based_approximate_Hxc_ener}
E^{(i)}_{\rm Hxc}(n_i)&=\innerop{\Psi^{(i)}(n_i)}{\hat{P}^{(i)}\hat{H}\hat{P}^{(i)}}{\Psi^{(i)}(n_i)}
\\
&\quad 
-\innerop{\Phi^{(i)}(n_i)}{\hat{P}^{(i)}\hat{h}\hat{P}^{(i)}}{\Phi^{(i)}(n_i)}
.
\end{split}\label{eq:approx_local_Hxc_dens_func_ener_def}
\ee
Indeed, $\Psi^{(i)}(n_i)$ and $\Phi^{(i)}(n_i)$ sharing the same core electrons (see Eq.~(\ref{eq:projector_op_onto_embedding_cluster})), it comes from the local density constraints of Eqs.~(\ref{eq:local_int_dens_constraint}) and (\ref{eq:local_nonint_dens_constraint}), by analogy with the rewriting of Eq.~(\ref{eq:KS_cluster_Hamil_full_proj}) into Eq.~(\ref{eq:non_int_KS_cluster_dens_func_Hamil_all_on_imp}), that  
\be\label{eq:proof_adding_Vext_into_local_Hxc_exp}
\begin{split}
&\innerop{\Psi^{(i)}(n_i)}{\hat{P}^{(i)}\hat{V}^{\rm ext}\hat{P}^{(i)}}{\Psi^{(i)}(n_i)}
\\
&
-\innerop{\Phi^{(i)}(n_i)}{\hat{P}^{(i)}\hat{V}^{\rm ext}\hat{P}^{(i)}}{\Phi^{(i)}(n_i)}
\\
&=
\left(v^{\rm ext}_i-\sum_{k}\inner{b^{(i)}}{\chi_k}^2v^{\rm ext}_k\right)
\\
&\quad
\times\left(\innerop{\Psi^{(i)}(n_i)}{\hat{n}_i}{\Psi^{(i)}(n_i)}-\innerop{\Phi^{(i)}(n_i)}{\hat{n}_i}{\Phi^{(i)}(n_i)}\right)
\\
&=0.
\end{split}
\ee
If, for convenience, we rewrite Eq.~(\ref{eq:cluster_based_approximate_Hxc_ener}) as follows,
\be
\begin{split}
E^{(i)}_{\rm Hxc}(n_i)&=\innerop{\Psi^{(i)}(n_i)}{\hat{P}^{(i)}\hat{H}\hat{P}^{(i)}}{\Psi^{(i)}(n_i)}
\\
&\quad 
-\innerop{\Phi^{(i)}(n_i)}{\hat{P}^{(i)}\hat{h}^{\rm KS}\hat{P}^{(i)}}{\Phi^{(i)}(n_i)}
\\
&\quad
+
\innerop{\Phi^{(i)}(n_i)}{\hat{P}^{(i)}\left(\sum_k v_k^{\rm Hxc}\hat{n}_k\right)\hat{P}^{(i)}}{\Phi^{(i)}(n_i)}
,
\end{split}
\ee
we obtain the final simplified expression (see Eqs.~(\ref{eq:local_nonint_dens_constraint}) and (\ref{eq:proof_adding_Vext_into_local_Hxc_exp})),
\be
\begin{split}
E^{(i)}_{\rm Hxc}(n_i)&=\innerop{\Psi^{(i)}(n_i)}{\hat{P}^{(i)}\hat{H}\hat{P}^{(i)}}{\Psi^{(i)}(n_i)}
\\
&\quad 
-\innerop{\Phi^{(i)}(n_i)}{\hat{P}^{(i)}\hat{h}^{\rm KS}\hat{P}^{(i)}}{\Phi^{(i)}(n_i)}
\\
&\quad+
\left(v^{\rm Hxc}_i-\sum_{k}\inner{b^{(i)}}{\chi_k}^2v^{\rm Hxc}_k\right)n_i
\\
&\quad
+2 \sum_{k}\inner{b^{(i)}}{\chi_k}^2v^{\rm Hxc}_k
,
\end{split}
\ee
where the last term simply originates from the fact that the embedding cluster contains (in the present case) two electrons. Note that, by construction, this term does not vary with the local density $n_i$. As a result, we deduce from Eqs.~(\ref{eq:local_dens_func_int_cluster_Hamil}) and (\ref{eq:local_dens_func_KS_cluster_Hamil}) that
\be
\begin{split}
\dfrac{\partial E^{(i)}_{\rm Hxc}(n_i)}{\partial n_i}&=\mu^{(i)}_{\rm imp}(n_i)-\mu^{(i)}_{\rm s}(n_i)
\\
&\quad
+
\left(v^{\rm Hxc}_i-\sum_{k}\inner{b^{(i)}}{\chi_k}^2v^{\rm Hxc}_k\right).
\end{split}
\ee
By applying the above equation to the special case where the local density equals the exact one, and combining the resulting expression with the local approximation of Eq.~(\ref{eq:local_approx_within_cluster}), we immediately recover from Eq.~(\ref{eq:no_imp_chem_pot_KS_system_exact_dens}) the LPFET impurity chemical potential expression that has been introduced in Eq.~(\ref{eq:final_LPFET_exp_chem_pots_wrt_Hxc_pot}) and obtained as an approximation to the exact theory. This alternative derivation confirms the physical relevance of LPFET in strongly correlated regimes.  




\newcommand{\Aa}[0]{Aa}
%


\end{document}